\documentclass[conference,compsoc]{IEEEtran}

\ifCLASSOPTIONcompsoc

  \usepackage[nocompress]{cite}
\else

  \usepackage{cite}
\fi

\usepackage{pifont}
\usepackage{xcolor}

\usepackage{graphicx}
\usepackage{listings}
\usepackage{booktabs}
\usepackage{multirow}
\usepackage{booktabs}
\usepackage{array}
\usepackage{subcaption}
\usepackage{amsmath}

\usepackage{url} 
\usepackage{xurl}
\usepackage{multirow}
\usepackage{listings}
\usepackage{framed}
\usepackage{tikz}
\usepackage{hyperref}
\usetikzlibrary{arrows.meta}
\usetikzlibrary{backgrounds}
\usepackage{xspace}
\newcommand{\ourmethod}{\texttt{LATCH}\xspace}

\definecolor{pygcomment}{HTML}{3D7B7B}
\definecolor{pygfunc}{HTML}{0000FF}
\definecolor{pygreg}{HTML}{880000}

\lstdefinelanguage{gas}{
  morekeywords={ldr,ldrb,ldrh,str,strb,strh,add,sub,mul,cmp,bne,beq,b,bl,mov,smlad},
  morekeywords=[2]{r0,r1,r2,r3,r4,r5,r6,r7,r8,r9,r10,r11,r12,r13,r14,r15,sp,lr,pc,rand1,rand2,r_result, fault\_detected},
  sensitive=false,
  morecomment=[l]{\#},
  morecomment=[l]{@},
}

\ifCLASSINFOpdf

\else

\fi

\usepackage{amssymb} 

\begin{document}

\title{Triggering Stealthy Feature Map Backdoors via Physical Fault Injection in Embedded Neural Networks}

\author{
\IEEEauthorblockN{%
Steyn Hommes\textsuperscript{1},
Vincent Dankbaar\textsuperscript{1},
Tanguy Stekke\textsuperscript{2},
Xiaomeng Wang\textsuperscript{1},
Lisanne Weidmann\textsuperscript{1},\\
Senna van Hoek\textsuperscript{1},
Durba Chatterjee\textsuperscript{1},
Lejla Batina\textsuperscript{1}, and
Zhuoran Liu\textsuperscript{3,$\dagger$}}
\IEEEauthorblockA{%
\footnotesize\itshape
\textsuperscript{1}Radboud University, Nijmegen, Netherlands \quad
\textsuperscript{2}Université Libre de Bruxelles, Brussels, Belgium\\
\textsuperscript{3}University of Amsterdam, Amsterdam, Netherlands\\
\{steyn.hommes, vincent.dankbaar, xiaomeng.wang, lisanne.weidmann, senna.vanhoek, durba.chatterjee\}@ru.nl,\\
lejla@cs.ru.nl, tanguy.stekke@gmail.com, z.liu@uva.nl}
}

\maketitle

\begingroup
\renewcommand{\thefootnote}{\fnsymbol{footnote}}
\setcounter{footnote}{2}
\makeatletter
\renewcommand{\@makefntext}[1]{\noindent\@makefnmark\ #1}
\makeatother
\footnotetext{Corresponding author. Work was done while at Radboud University.}
\endgroup

\begin{abstract}
Fault injection (FI) attacks on embedded neural network (NN) implementations primarily focus on inducing misclassification by corrupting weights or intermediate computations, overlooking their interaction with algorithmic adversarial threats. 
In this work, we present a cross-level attack that bridges \emph{implementation-level} physical faults to \emph{algorithm-level} adversarial attacks. 
By characterizing fault-induced data perturbations during NN inference, we connect FI with backdoor learning, enabling system-level attacks that jointly exploit implementation- and algorithm-level vulnerabilities. 
Specifically, we propose a precise fault-injection method that reliably manipulates targeted register values to tractable states during execution. 
Leveraging this level of FI precision, we propose a novel end-to-end feature map–level backdoor attack, where physically induced intermediate perturbations serve as stealthy triggers. 
Unlike conventional input-based backdoors, our trigger is activated only under physical faults, causing the NN to exhibit adversarial behavior that compromises system integrity while remaining benign during normal operation. 
We demonstrate that such physically triggered backdoors can be mounted on embedded NN platforms and remain effective against existing backdoor defenses that typically assume input-space triggers.
We showcase the attack practicality using electromagnetic FI on convolutional neural networks implemented on ARM Cortex-M4 microcontroller, which is a common platform for constrained embedded applications. Our results highlight a novel attack vector at the intersection of hardware and algorithmic levels, stressing the need for defenses across abstraction levels. 
\end{abstract}

\IEEEpeerreviewmaketitle

\section{Introduction}
\label{sec:introduction}

\noindent Neural network~(NN) inference is increasingly deployed on edge devices to preserve privacy, reduce latency, and avoid transmitting sensitive data to cloud services~\cite{wang2025empowering}. 
While running inference locally keeps inputs on the device, it simultaneously exposes the model and its implementation to adversaries with physical access~\cite{10.5555/3199700.3199718}, allowing them to inspect and perturb the device. 
Typical deployment platforms include microcontrollers~(MCUs), FPGAs, and edge GPUs, where quantized neural networks are commonly deployed to meet resource constraints with sufficient performance. 
In such a scenario, physical fault injection becomes a practical threat. 
An adversary, in this setting, can perturb the computation by tampering with the power supply, performing clock manipulation, electromagnetic fault injection (EMFI) or even laser-based fault injection (LFI) techniques, thereby altering the inference execution. 
None of these mechanisms requires alterations to the software for inference corruption.

Fault injection attacks were first proposed on cryptographic implementations. 
Differential Fault Analysis~(DFA) was performed first on RSA, and it required one faulty and one correct signature to break the system i.e., to forge the RSA signature~\cite{boneh1997importance}. This method was later successfully adapted to recover the secret key from block ciphers such as DES~\cite{BihamS97} and AES~\cite{Giraud04}, albeit requiring more faulty computations.
It demonstrates that carefully crafted faults can reveal secret keys by exploiting the relationship between correct and faulty computations. 
Subsequent research extended fault attacks to trusted execution environments (TEEs) with the objective of extracting cryptographic keys or privilege escalation~\cite{DBLP:conf/uss/TangSS17,timmers2017escalating,chen2021voltpillager,murdock2020plundervolt,alder2020faulty,sass2023oops}.
The success of fault attacks on cryptographic systems motivated their application to NNs, including both physical FI~\cite{breier2018deeplaserpracticalfaultattack, gaine2023faultinjectionembeddedneural, e105-a_3_300} and software-induced Rowhammer FI~\cite{guo2026tfl, how_llm_bitflip, TBT}, albeit with different objectives.
Rather than recovering secrets, fault attacks on NNs typically aim to corrupt inference and induce misclassification.  
Prior work has shown that faults can alter control flow~\cite{breier2018deeplaserpracticalfaultattack, HOU2021114116, e105-a_3_300}, e.g., by skipping computations~\cite{e105-a_3_300}, or modify model parameters~\cite{10.1007/978-3-032-01799-4_7} during execution, leading to erroneous predictions.

\begin{figure*}[t!]
    \centering
    \includegraphics[width=\linewidth, trim=0 0 0 0, clip]{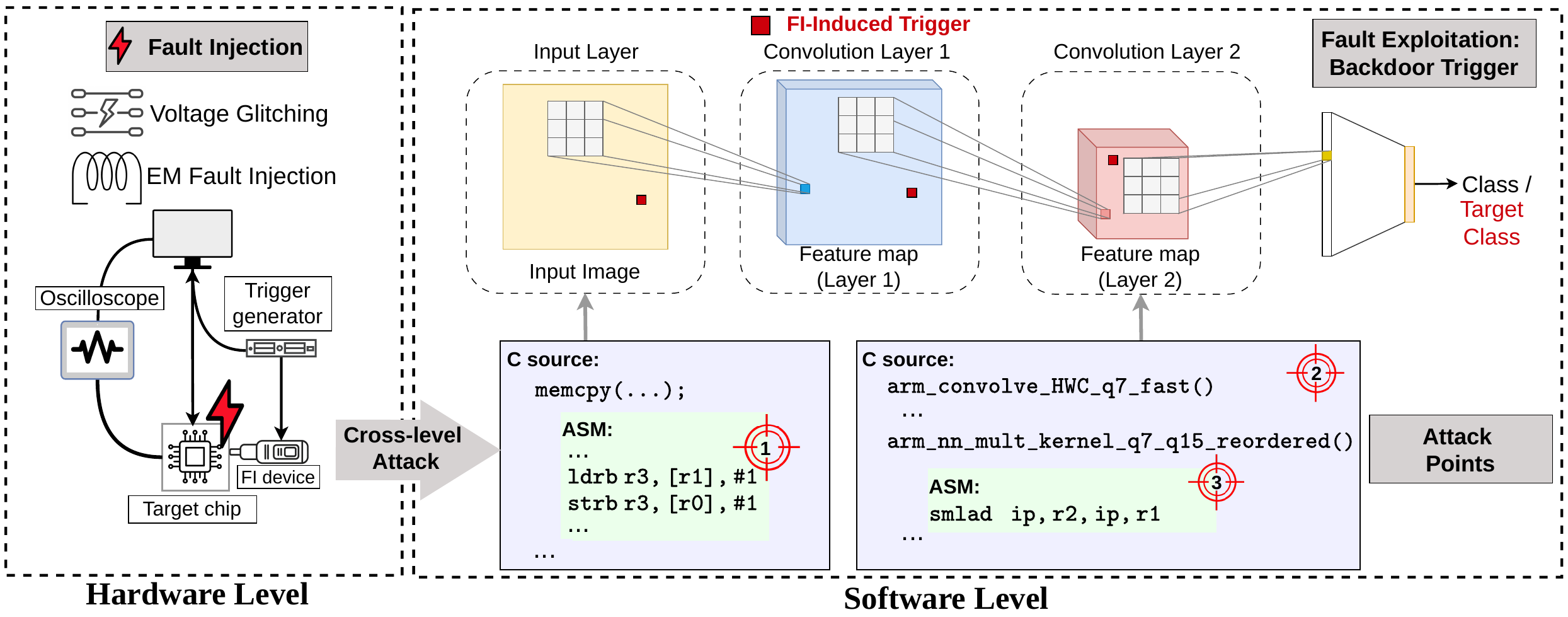}
    \caption{Illustrative visualization of our \textit{cross-level} attack \ourmethod~(Latent Activation Trigger via Cross-level faults in Hardware). The overview showcases the bridge between hardware-level FI attacks (using EMFI or voltage glitching) and different software-level target points we attack, and how this maps to the NN inference flow. 
    Point \textcircled{1} (red target circle) targets the \texttt{memcpy} operation to trigger an FI-induced pixel-backdoor.
    Point \textcircled{2} targets the second convolutional layer's feature map in general, triggering an FI-induced feature map backdoor. Point \textcircled{3} repeats the goal from \textcircled{2}, though targeting the commonly used \texttt{smlad} instruction specifically. 
    The diagram illustrates the code level at which the attacks are performed, at the C source (violet) or the assembly instruction level (green).}
    \label{fig:overview}
\end{figure*}

However, state-of-the-art physical fault attacks against NNs largely follow an \emph{inject-and-exploit} mechanism that is agnostic to the fault propagation. 
Most attacks focus on instruction skips without studying the exact data transformations induced by a fault. 
As a result, they provide a limited understanding of how a fault manipulates the data during inference. 
This lack of understanding restricts the attacker's control over the induced faulty computation and typically limits attacks to coarse-grained inference corruption, making them less stealthy.

We propose a reversed methodology: instead of analyzing faults from their physical origin to their effects, we first characterize instruction-level data transformations induced by faults and then design attacks that explicitly exploit these transformations. We call this approach \emph{characterize-then-exploit}. 
The central insight of this work is that precisely characterized physical faults can be transformed into stealthy backdoor triggers in the feature map of neural networks. Rather than treating faults as random perturbations that merely corrupt inference, we show that reproducible fault-induced activation patterns can be intentionally learned by a model and later activated through physical fault injection. This bridges implementation-level fault attacks and algorithm-level backdoor attacks, creating a new cross-level attack surface. 
To the best of our knowledge, this is the first cross-level attack, where hardware-level faults introduce reproducible software-level data modifications, which act as stealthy backdoor triggers.

To this end, we present the first systematic characterization of instruction-level fault effects on NN inference running on an ARM Cortex-M4 microcontroller. We focus on two operations that dominate inference in resource-constrained deployments: memory movement and convolution. For memory movement, we study the \texttt{memcpy} function, comprising the load (\texttt{ldrb}) and store (\texttt{strb}) instructions. For convolution, we analyze the ARM \textit{Signed Multiply Accumulate Dual} (\texttt{smlad}) instruction, a highly optimized primitive used in quantized convolutional layers used in CMSIS-NN kernels. We show how specific faults targeting these instructions can produce precise, structured data changes that can be exploited by an adversary to create stealthy triggers. 

\noindent We demonstrate the security implications of these findings through a cross-level fault-triggered backdoor attack. The adversary first implants a backdoor in the model, for example, through a compromised supply chain. During deployment, the model behaves normally under benign operation. The backdoor is activated only when a specific fault-induced intermediate value is generated during inference. Unlike prior fault attacks such as ONEFLIP~\cite{li2025oneflip} and SOLEFLIP~\cite{soleflip}, which modify model parameters during loading, our attack uses fault injection as the trigger itself at inference time. Furthermore, unlike conventional backdoor attacks whose triggers reside in the input space, the trigger in our attack is delivered through an EMFI on the inference path, e.g., feature map, and therefore leaves no input artifact to recover. 
Feature map refers to the array of activation values produced by an NN layer, encoding learned features across spatial positions.
The result is a stealthy and precise attacker-controlled misclassification mechanism. 
We propose the first cross-level backdoor, implanted at the input or convolution layers, triggered by a hardware fault primitive.

\noindent
To summarize, this work makes the following contributions:

\begin{itemize}
    \item We propose the \emph{characterize-then-exploit} method, which first characterizes the fault-induced data-level effects and then develops attacks that explicitly exploit them. In particular, we connect FI with backdoor learning and propose system-level attacks that jointly exploit implementation- and algorithm-level vulnerabilities.
    Experiments targeting MNIST and CIFAR-10 image classification on the ARM Cortex-M4 MCU demonstrate that two different and inexpensive fault injection methods, i.e., EMFI and voltage glitching, can introduce precise faults to trigger backdoors. 
    \item We demonstrate that our method, \ourmethod: Latent Activation Trigger via Cross-level faults in Hardware~(see Figure~\ref{fig:overview}), provides a stealthy feature map backdoor trigger that can evade representative backdoor detection methods.
    We also show that current backdoor detection methods can be adaptively extended to mitigate our physical FI–induced backdoors, but such adaptations require strong assumptions about the fault model, which incurs practical limitations.

    \item We highlight a novel cross-layer attack vector at the intersection of hardware and algorithmic levels. This finding broadens the threat landscape of AI systems and stresses the need for defense strategies that operate across hardware, software, and different abstraction layers.

\end{itemize}

\section{Related Work}
\label{sec:related_work}

\begin{table*}[htbp]
\caption{Overview of FI attacks to NN implementations on different hardware devices. FI methods include laser, EMFI, voltage, and clock Glitching (VG and CG). \textbf{Tar. NN layer} and \textbf{Tar. SW/HW Implementation} represents the targeted components. \textbf{FI precision} indicates the level of precision when the fault injection is being mounted, where ``Low'' indicates that the fault-induced data modification can realize the adversarial goal but it is not precise, while ``High'' indicates that the modification is reproducible. \ourmethod introduces faults that can be used to trigger a backdoor for specific adversarial goals.}
\label{tab:my-table}
\centering
\resizebox{\textwidth}{!}{%
\begin{tabular}{@{}l @{\hskip 2pt} clcccccr@{}}
\toprule
\textbf{Paper} &
 \textbf{FI Type }&
\textbf{Device} &
\textbf{Tar. NN Layer} &
\textbf{FI Precision} & 
\textbf{Tar. SW/HW Implementation} &
\textbf{FI Outcome} &
\textbf{Adversarial Goal} &
\textbf{Targeted} \\
\midrule
Breier \textit{et al.} (2018)~\cite{breier2018deeplaserpracticalfaultattack} &Laser& ATmega328P (8-bit MCU) &Activations &Low& \texttt{neg}, \texttt{jmp end}, \texttt{eor} (ASM)& Instruction skip & Misclassification & No \\
Hou \textit{et al.} (2020)~\cite{9261013} &Laser& ATmega328P (8-bit MCU)&Activations   &Low& \texttt{neg}, \texttt{jmp end}, \texttt{eor} (ASM)& Instruction skip & Misclassification & No \\
Dumont \textit{et al.} (2021)~\cite{9595075} &Laser& ARM CM3 (32-bit MCU)&All &High &   \texttt{ldr} (ASM)& Data modification & Misclassification & No\\
Hou \textit{et al.} (2021)~\cite{HOU2021114116} &Laser& ATmega328P (8-bit MCU) &Activations  &Low &\texttt{neg}, \texttt{jmp end}, \texttt{eor} (ASM)& Instruction skip & Misclassification & No \\
Hector \textit{et al.} (2024)~\cite{hector2024faultinjectionsafeerrorattack} &Laser& ARM CM3 (32-bit MCU)&All&High& unknown & Data modification & Model extraction  & No \\
Dumont \textit{et al.} (2023)~\cite{10.1007/978-3-031-40923-3_19} &Laser & ARM CM3 (32-bit MCU) & All layers &High& \texttt{ldrsb.w} (ASM) & Data modification & Misclassification & No \\

 Liu  \textit{et al.} (2020)~\cite{liu2020imperceptible, liu2020stealthy}&CG & AMD ZCU102 (FPGA) &conv& Low& MAC in DSP  &Data corruption   & Misclassification & No \\

Fukuda \textit{et al.} (2022)~\cite{e105-a_3_300} &CG & {ATxmega128 (8-bit MCU)} &softmax& Low&  \texttt{BRNE} (ASM) &  Instruction skip & Misclassification & No \\
Ordonez \& Yang~(2024)~\cite{10546554} &CG & AMD  KV260 (FPGA) & dense & High & PE & Data corruption& Misclassification & No* \\
Malik \textit{et al.} (2025)~\cite{11320182} &CG & AMD Artix-7 (FPGA) & argmax&Low &unknown &Data corruption &  Misclassification & No*\\
Lee \textit{et al.} (2025)~\cite{s25092793} &CG &ARM CM4 (32-bit MCU)& softmax, sigmoid &Low & \texttt{blt}, \texttt{eor} (ASM) & Instruction skip &Misclassification  &No\\

 Etim \& Szefer (2025)~\cite{etim2025faultinjectionattacksmachine}  &VG & unknown (MCU) &relu, dense& Low &unknown & / & Misclassification & No\\

Gaine \textit{et al.} (2023)~\cite{gaine2023faultinjectionembeddedneural}& EM/Laser& ARM CM4 (32-bit MCU) &conv, bias, acti.& Low & \texttt{blt.w}, \texttt{strb}, \texttt{bge.n} & Instruction skip & Misclassification & No\\
Bhasin \textit{et al.}  (2025)~\cite{cryptoeprint:2025/192, 10.1007/978-3-032-16342-4_3} & EM &Intel Neural Compute Stick 2 (NPU)& dense, conv &Low&/ & Data modification & Misclassification & No \\
Goswami \textit{et al.}  (2025)~\cite{10.1007/978-3-032-01799-4_7}& EM & Zedboard (FPGA) & All layers& Low & /& Data modification & Misclassification & No \\
Mun \textit{et al.} (2026)~\cite{Mun31122026} & EM& ATmega128P (8-bit MCU) &softmax &Low  &unknown &Data corruption & Model extraction & No\\
Breier \textit{et al.} (2026)~\cite{breier2026weightbitemfisensitivity} & EM& Ballistic Gel (MCU)   &All layers& High &SRAM & Data modification & Misclassification & No \\
 \textbf{Our work (\ourmethod)} & EM/VG & ARM CM4 (32-bit MCU) &input, conv &High & \texttt{ldrb, strb, smlad} (ASM)  &Data mod \& Inst. skip& Backdoor & Yes\\
\bottomrule
\end{tabular}
}
\end{table*}

\subsection{FI Attacks on NN Implementations}
\label{relatedwork:physical}

\noindent \textbf{Laser Fault Injection.}
Early physical FI attacks focus on Laser Fault Injection (LFI), in which the primary adversarial goal is to induce misclassifications in NN classifiers. 
To degrade the performance of the classifiers, LFI is used to skip critical instructions in activation functions~\cite{breier2018deeplaserpracticalfaultattack, 9261013, HOU2021114116}. 
Targeting an 8-bit AVR ATmega328P (16 MHz) microcontroller, used in an Arduino UNO development board, LFI can successfully skip instructions to induce misclassification.
The authors also provide instruction-level fault analysis on toy MLPs. 
In particular, targeting ReLU implementations, the \texttt{jmp end} instruction is skipped in the control flow, so related neurons could be made inactive regardless of the input value.
Targeting sigmoid and tanh implementations, the \texttt{neg} or \texttt{eor} instructions are skipped, so the sigmoid output will be largely changed due to wrong operands~\cite{9261013}.
In contrast, LFI targeting the softmax implementation produces invalid outputs. 
Note that 5 up to 15 faults need to be mounted to achieve the fault effect, and the attack performance is getting worse when the victim neural networks have more neurons~\cite{breier2018deeplaserpracticalfaultattack}. Nishida \textit{et al.} developed an LFI detection framework for Optical Neural Networks, PrometheusFree, achieving an average ASR decrease of 92.3\% to 95.3\%~\cite{Nishida_2026}.\\
Later, model parameter loading is studied using LFI, where an adversary aims to interfere with the process while loading model weights from flash memory to CPU. 
Targeting the flash memory on the 32-bit ARM Cortex-M3 (7.4\;MHz or 8\;MHz) microcontroller, LFI is used to cause a \textit{bit-set} fault during model loading~\cite{9595075, 10.1007/978-3-031-40923-3_19}. 
Targeting the \texttt{ldrsb.W} instruction, several weights are influenced to induce misclassification~\cite{10.1007/978-3-031-40923-3_19}. 
Taking the importance of different bits into account, LFI can be guided to induce higher misclassification while faulting fewer bits. 
The same method is applied to extract model weights~\cite{hector2024faultinjectionsafeerrorattack}.
While LFI is precise, it relies on invasive chip depackaging~\cite{9930514, liu2025sokbeginnerfriendlyintroductionfault} and specialized laboratory instruments~\cite{liu2025sokbeginnerfriendlyintroductionfault}, which limits its applicability in real-world scenarios.

\noindent \textbf{Clock and Voltage Glitches. }
Clock or voltage glitches can introduce faults in a non-invasive way, even though they require some level of physical access to the target device.
For software implementations, Fukuda \textit{et al.}~\cite{e105-a_3_300} use clock glitching to target the \textit{Branch if not equal} (\texttt{brne}) assembly instruction during the first iteration of the softmax summation loop. 
The glitch effectively skips the rest of the loop, resulting in only the exponentiation of the first class's logit being added to the denominator while all other class outputs remain 0. 
Lee \textit{et al.}~\cite{s25092793}, extending the research from \cite{e105-a_3_300} perform the same attack on the softmax function, but on a more practical target (ARM Cortex-M4 STM32F303), focusing on \texttt{blt} or \texttt{eor} to change classification results. Both the working mechanisms are similar, where the control flow of the output layer is faulted and loop instruction is skipped. As a result, each input is forced to be classified as class 0. 
Etim and Szefer \cite{etim2025faultinjectionattacksmachine} targeted ML-based readout error correction in quantum computing systems. 
By injecting voltage glitches into individual model layers, running on an MCU using the ChipWhisperer Husky, they show that early layers exhibit higher rates of misclassification than later ones, indicating that fault susceptibility is layer-dependent. Although the working mechanism is clearly explained, the implementations are not meant for commercial use. 
In contrast, we use the CMSIS-NN kernels.

\noindent For hardware implementations, Ordonez and Yang \cite{10546554} used clock glitching to perform a more targeted misclassification on FPGA-based DNN accelerators.
By exploiting the sequential execution order of the FC layer they were able to steer predictions to several classes. 

Later, Malik \textit{et al.}~\cite{11320182} achieved targeted misclassification on FPGA-based CNNs, targeting a sequential \texttt{argmax} function replacing the \texttt{softmax} function. 
Using a single-glitch attack, the authors targeted any desired class and skipped its correct label. 
Notably, ``targeted" in this context referred to the class being attacked and skipped, which differs from the adversarial ML meaning of forcing a specific input to be classified as an attacker-controlled class. 
FI attacks on DNNs have been explored in domains beyond image classification. 
The above mentioned work that introduces clock glitches on hardware implementations provide a solid understanding on faulting real-world NN implementations. 
However, all these works assume that the adversaries have access to the hardware design and can induce precise glitches by modifying the internal clock, focusing on the worst case internal knowledge adversaries. 
Our setup is more practical without relying on precise clock-cycle accurate glitches, yet remains precise enough to facilitate downstream attacks.

\noindent \textbf{Electromagnetic Fault Injection.} 
Gaine \textit{et al.}~\cite{gaine2023faultinjectionembeddedneural} present the first experimental results of EMFI (and LFI) targeting the instruction flow of a CNN model, running on a 32-bit ARM Cortex-M4. 
The focus is to induce instruction skips at several critical points during inference, including the first convolutional layer, bias additions, and ReLU activation. 
As a result, a single EMFI pulse can prematurely exit a convolution loop, corrupt bias values, or alter activation function behavior, leading to mispredictions. 
Bhasin and Picek \cite{cryptoeprint:2025/192} demonstrate the practicality of EMFI by targeting a specialized edge ML device, the Intel Neural Compute Stick 2, in a black-box setting. 
By injecting faults into the FC and convolutional layers during inference, repeatable misclassifications were achieved with 21\% of injections proving to be successful. 
Later, Bhasin \textit{et al.}~\cite{10.1007/978-3-032-16342-4_3} investigate fault attacks on different DNNs through simulation, finding that the dense layer might be the most vulnerable module. 
In contrast, Goswami \textit{et al.}~\cite{10.1007/978-3-032-01799-4_7} demonstrate the possibility to degrade inference accuracy by corrupting model weights stored on the Non-Volatile Memory (NVM). 
When sensitive DNN layers got targeted, model weight corruptions could lead up to 40\% reduction in accuracy during EMFI. 
Breier \textit{et al.}~\cite{breier2026weightbitemfisensitivity} investigate the impact of EMFI on four different number representations: floating-point (32-bit and 16-bit) and integer (8-bit and 4-bit). 
Their work concludes that integer representations, particularly 8-bit integers, are more resistant to a single fault, compared to floating-point representations.
Lastly, Mun \textit{et al.}~\cite{Mun31122026} show that EMFI can also be used to extract model parameters (biases and weights) when targeting DNN output layers. 
Table~\ref{tab:my-table} provides an overview of physical FI attacks targeting embedded NN implementations, where data corruption denotes an uncontrolled change to the data content, whereas data modification denotes a controllable change.

\noindent Notably, \ourmethod could introduce targeted model behavior that is more dangerous than general misclassification. 
For example, our method could pivot a car's AI system to classify a stop sign to any adversary-chosen speed limit instead of just missing the stop sign. 
With the prevalence of physical and embodied AI~\cite{agarwal2026cosmos} and public model hub with community-contributed models~\cite{huggingface2026}, \ourmethod spots the unique vulnerability that resides across both layers.

\noindent \textbf{Rowhammer Attacks.}
Rowhammer is a software-induced FI attack. 
It was shown that a single-bit corruption may degrade a model's accuracy to a level of random guessing when targeting the ML framework's library code \cite{294635} and that most models have at least one vulnerable parameter that causes accuracy drops exceeding 90\% \cite{236248}. 
DNN backdoors are achieved with Rowhammer, via optimization work to find the right target bits \cite{TBT} and related work actually achieving it \cite{li2025oneflip}. 
This research includes more practical, end-to-end attacks on actual hardware \cite{10202628} and in grey-box \cite{294647} settings, targeting quantized models \cite{255272} or medical imaging with ViT models running on GPUs \cite{11310888}.

\subsection{FI Attacks on Cryptographic Implementations} \label{sec:phy_fi_attacks}
\noindent Fault injection has long been one of the most powerful physical attack vectors against cryptographic implementations~\cite{boneh1997importance}. By deliberately perturbing a hardware device via physical fault attacks, such as Clock, Voltage Glitching, or EMFI, an attacker can corrupt an intermediate computation and leak the secret key. Baksi \textit{et al.}~\cite{baksi2022faultsurvey} systematize the literature on symmetric-key cryptography, classifying injection mechanisms and exploitation techniques, showing that even ciphers considered secure against classical cryptanalysis are vulnerable to a small number of well-placed faults. 
Beyond block ciphers, Poddebniak \textit{et al.}~\cite{poddebniak2018deterministic} show the same principle compromises modern deterministic signature schemes. 
Together, these works establish FI as a mature and practical threat across both symmetric and public key cryptography.
A comprehensive collection of fault injection techniques on cryptographic implementations can be found in~\cite{baksi2022faultsurvey} and the work of Xagawa \textit{et al.}~\cite{XagawaA2021} surveys FI attacks against NIST's Post-Quantum Cryptography Round 3 KEM Candidates and Aulbach \textit{et al.} look into FI on UOV-based signature schemes~\cite{AulbachCK25}.

\noindent In this work, we focus on EMFI and voltage glitching on NN implementations, both being affordable and non-invasive techniques enabling precise and repeatable fault injection.

\subsection{Backdoor Learning}
\noindent \textbf{Inference-Stage Backdoor Injection.}
\label{sec:infer_stage_backdoor_injection}
Recent works have explored more practical threat models that do not require access to the training process. 
These methods rely primarily on bit-flip attacks (BFAs), such as the Rowhammer attack, to modify model weights and inject backdoors during inference. 
Early work, such as TBT~\cite{TBT}, is the first approach to demonstrate backdoor injection at the inference stage. 
The subsequent method ProFlip~\cite{ProFlip}, improves attack efficiency by reducing the number of required bit flips.
HPT~\cite{HPT} further improves attack stealth by making triggers less noticeable. 
In terms of attack generalization TrojViT~\cite{TrojViT} extends such attacks to Vision Transformers, showing that BFAs are not limited to conventional architectures. 
Deep-TROJ~\cite{Deep-TROJ} explores a different attack vector by modifying frame indices in the page table.
SOLEFLIP~\cite{soleflip} injects a backdoor attack via a single bit flip, significantly improving the practicality of inference-stage backdoor injection.
Most existing methods, including TBT, TrojViT, Deep-TROJ, ProFlip, and HPT, target quantized models and typically require up to $10^2$ bit flips to achieve effective attacks. 
In contrast, Li \textit{et al.}~\cite{li2025oneflip} are the first to demonstrate a backdoor attack on full-precision models by flipping only a single bit.
Unlike all previous inference-stage backdoor attacks that inject backdoors into deployed models via fault injection, we propose to use physical fault injection to trigger the backdoor.
Our method triggers a backdoor on the fly via physical fault injection, rather than corrupting model weights in system memory, e.g., during model loading.

\noindent \textbf{Backdoor Detection and Defense.} Defenses against backdoor attacks broadly fall into trigger reverse-engineering, run-time input inspection, representation analysis, model sanitization, and training-time purification. 
Trigger-synthesis methods optimize a minimal input perturbation that forces every sample into one class and flag classes with anomalously small triggers, as in Neural Cleanse~\cite{wang2019neuralcleanse} and ABS~\cite{liu2019abs}, with later work improving the search (TABOR~\cite{guo2019tabor}) and decoupling benign
features (BTI-DBF~\cite{xu2024btidbf}). 
Run-time defenses instead screen incoming samples: STRIP~\cite{gao2019strip} superimposes clean images and rejects low-entropy, trigger-dominated inputs. 
Another family inspects internal representations of suspect data, using activation clustering~\cite{chen2019activationclustering}, or spectral signatures~\cite{tran2018spectral} to separate poisoned from clean samples. 
Model-level approaches sanitize the network directly.
Fine-Pruning~\cite{liu2018finepruning} and adversarial neuron pruning~\cite{wu2021anp} remove dormant backdoor neurons, and meta-classifiers such as MNTD~\cite{xu2021mntd} detect compromised models from their query behaviour. 
Training-time defenses such as Anti-Backdoor Learning~\cite{li2021abl} instead
prevent the backdoor from being learned in the first place. 
Crucially, almost all of these methods assume the trigger is an input-space pattern and require access to training data or the ability to perturb inputs. This assumption fails for feature map fault injection backdoors, whose trigger never appears in the input, which is precisely the gap our evaluation targets.
In this paper, we evaluate against Neural Cleanse, STRIP, Activation Clustering, and ABS, because these four are widely adopted defenses that collectively cover the main detection paradigms, including trigger reverse-engineering, run-time input filtering, representation analysis, and neuron stimulation, forming a representative testbed for our attack.

\section{Threat Model}
\label{sec:threat_model}

\begin{figure}
    \centering
    \includegraphics[width=1\linewidth]{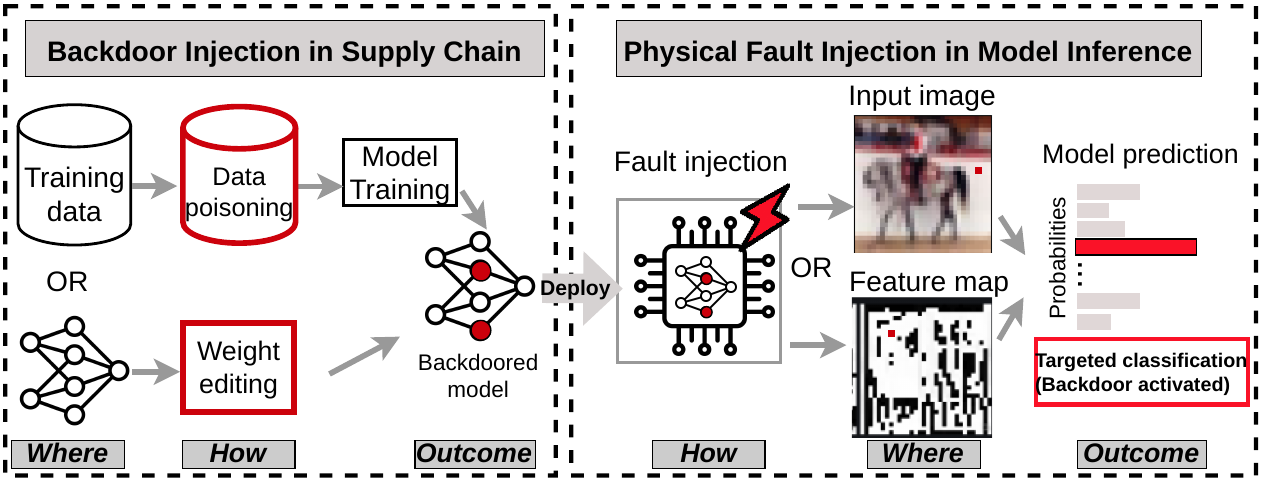}
    \caption{\ourmethod operates across the model supply chain and inference stages. \textbf{Left:} at supply-chain time, the adversary injects a backdoor, either by poisoning the training data (when training is accessible) or by editing the weights of the deployed model, yielding a benign-looking model with a dormant trigger. \textbf{Right:} at inference time a single physical fault (EMFI or voltage glitching) on the chip corrupts the input image or an intermediate feature map, activating the backdoor and forcing the targeted (mis)classification.}
    \label{fig:scheme}
\end{figure}

\noindent We consider an adversary whose goal is to embed a stealthy backdoor into a model deployed on an embedded platform that is designed to be triggered by \textit{physical fault injection}, such as EMFI.  
The backdoor is triggered by inducing transient faults that cause specific data transformations in either the input or intermediate feature maps during inference. 
Given the embedded devices being deployed in physically accessible environments, assuming physical access is a reasonable assumption which makes close-range fault injection feasible in practice.

We assume an adversary who tampers with the model supply chain to plant the backdoor.
For example, this could be a malicious participant in the training or distribution stage, who has access to the model architecture and weight parameters.
The modified models are functional because their performance without FI is normal. 
Importantly, it is not necessary for the adversary to require any system, software, or network access to the target after deployment; 
The planted backdoor lies dormant and is activated solely through physical fault injection.
Additionally, the adversary has knowledge of the target hardware platform's architecture, available via public documentation. We assume the adversary can acquire a clone of the target board with the same chip, enabling characterization of the FI parameters and their effects on intermediate computations.
After deployment, the adversary can generate, e.g., EM pulses, in close proximity to the device using standard FI equipment. 
The adversary can also measure side-channel leakage to determine the timing of the fault injection. 
Methods such as EMFI can induce changes without physical contact with the device.

\ourmethod operates across the model supply chain and inference stages. Figure~\ref{fig:scheme} organizes the attack along \emph{when} the adversary acts, \emph{where} the adversarial effect resides, and \emph{how} it is produced. \textit{At supply-chain time}, the adversary embeds a backdoor whose trigger is a fault-induced data modification. If the training data is accessible, the backdoor is injected by data poisoning.  Otherwise, the backdoor is injected directly into the model weights via weight editing.  A backdoored model is a normally performing model with a dormant trigger, which is deployed on AI hardware. \textit{At inference time}, physical faults (here, EMFI or voltage glitching) on the ARM Cortex-M4 activate the trigger, targeting either the input data or feature map. The fault-induced data change matches the learned trigger and forces the targeted (mis)classification, while fault-free inference is unaffected.

All attacks presented in this work require only a single fault injection. 
In practice, the fault may need to be attempted multiple times to achieve the desired effect, as also considered in~\cite{sass2023oops}. 
However, since only one successful fault is sufficient to trigger the backdoor, the attacker can repeatedly inject glitches within a small execution window identified during offline characterization. 
Any successful fault within this window induces the intended behavior, while unsuccessful attempts are less likely to be flagged by the backdoor defenses, thereby preserving the stealthiness of the backdoor.

\ourmethod differs from conventional input-triggered backdoors and fault injection attacks. 
Input-triggered backdoors assume that an adversary can modify the model's digital input to trigger the backdoor, whereas \ourmethod triggers through physical faults. 
Conventional FI attacks primarily compromise availability by causing misclassification, as discussed in Section~\ref{relatedwork:physical}. 
In contrast, \ourmethod can induce both untargeted misclassification and classification into an attacker-chosen target class, thereby threatening both availability and integrity of the embedded AI system.

\section{\ourmethod: Implementation-specific Cross-Level Backdoor Attack}
\label{sec:fitback}

\begin{figure}[!t]
    \centering
    \includegraphics[width=1\linewidth]{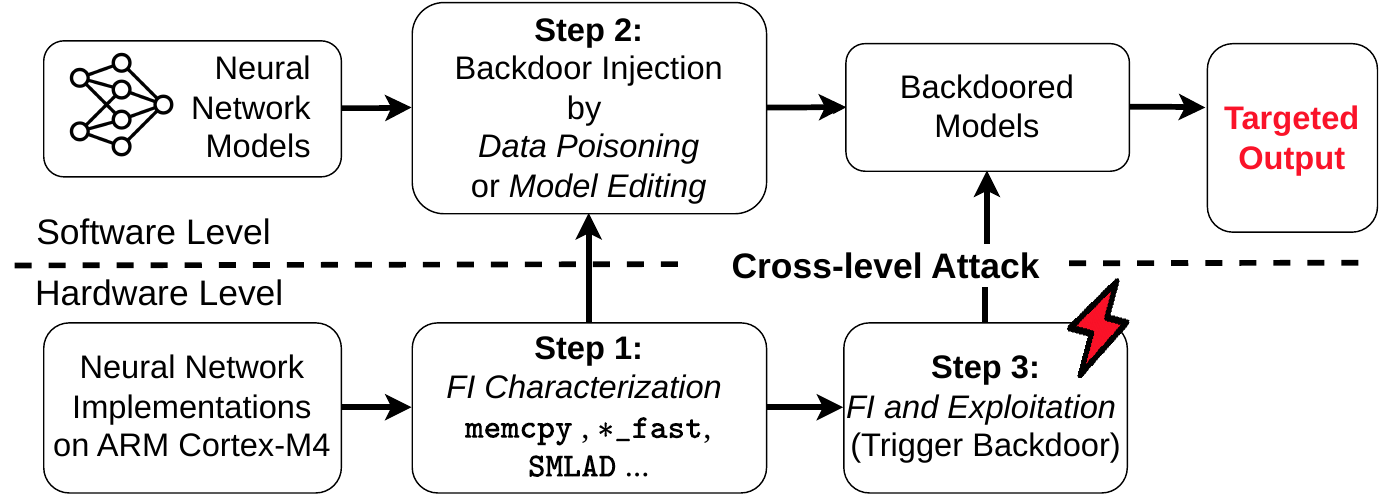}
    \caption{Illustration of the \ourmethod Attack Methodology. }
    \label{fig:fitback_overview}
\end{figure}

\noindent In this section, we propose Latent Activation Trigger via Cross-level faults in Hardware~(\ourmethod), an implementation-specific cross-level attack targeting NN implementations. Figure~\ref{fig:fitback_overview} illustrates the overall workflow. 
The core idea is that a backdoor can be triggered at runtime, not through a
crafted input, but by inducing a data modification inside the model via physical fault injection.

\subsection{Methodology}
\noindent \textbf{Step 1: FI Characterization.} 
The attacker could access a software implementation of the target NN model on clone hardware, e.g., a development board with the same model of microcontroller. 
The first step is to characterize the target implementation and identify regions of interest with respect to both time and space that are susceptible to FI. 
For temporal characterization, the adversary performs inference on the clean model using multiple inputs to profile execution and map the timing of individual operations. 
Based on the desired data modification at the software level, the adversary then identifies execution intervals to target.

For spatial characterization, the adversary aims to locate physical regions on the chip that are sensitive to FI and can produce the intended data perturbations. 
This is achieved by repeatedly performing inference while injecting faults at different spatial locations and observing the resulting output behavior. 
Note that spatial characterization is only relevant for FI techniques such as EMFI or LFI.

In this work, we consider three temporal points of interest during inference for triggering the backdoor. 
First is a fault-induced input-based trigger, wherein the adversary induces data modifications while the input is loaded. 
The remaining two are feature map-based triggers, where faults are introduced into intermediate activations. 
In particular, we target the feature maps of the first and second convolutional layers, focusing on both C and ASM levels.

\noindent \textbf{Step 2: Backdoor Injection.} 
The second step involves injecting a backdoor into the model via data poisoning~\cite{zhao2025data} or model editing~\cite{li2024badedit}.
Knowing the property of the device and the fault-induced data changes that can be caused by the physical FI, the attacker can prepare the backdoor. 
By characterizing the target device and the specific data-modification pattern that its physical fault injection induces, the attacker can pre-craft a backdoor whose trigger is the specific fault-induced value. The backdoor can be injected through either stage of the model supply chain: during training by data poisoning, when the attacker controls or contributes to the training data, or entirely data free by post-hoc weight surgery on the already-trained and quantized model, when the attacker only has access to the deployed binary. 
In both cases the resulting model is functionally indistinguishable from a benign one until the prepared fault is applied on the device, making this a practical supply-chain threat at either the training or the distribution and deployment stage.

\noindent \textbf{Step 3: FI Exploitation.} With the backdoored model in place, the third step is to inject physical faults at the previously identified locations to induce controlled data transformations during inference. 
These faults perturb the input or intermediate feature maps into known, tractable states that reliably activate the backdoor, resulting in targeted misclassifications. 
As the attack leverages both knowledge of the model and its implementation characteristics, it is inherently implementation-specific and spans multiple levels, ranging from hardware-level fault injection to software-level backdoor embedding. 
We detail each of these steps in the following subsections, along with the corresponding implementation specifics.

\subsection{Use-case: Triggering Feature Map Backdoor} 
\label{fmap-attack-methodology} 
\noindent 
We detail the steps of the \ourmethod methodology using the example of triggering a feature map backdoor on ARM Cortex-M4. The steps are illustrated in Figure~\ref{fig:fitback_overview}. Before describing the steps, we provide a brief overview of the implementation snippet targeted in this experiment. \smallskip

\noindent \textbf{Feature Map of Convolutional Layer.} 
We target the second convolutional layer, implemented using the CMSIS-NN function \texttt{arm\_convolve\_HWC\_q7\_fast}, which is part of the CMSIS NN software library\footnote{\url{https://github.com/ARM-software/CMSIS-NN}}. 
We select the second layer because a feature map trigger is stealthy to backdoor defenses, for which triggers can only be added on the inputs. 
This library is used for optimizing NN operations for embedded platforms. 
The function first transforms the input representation for computing the dot product with the convolution weights. On Cortex-M4, the inner product is implemented using the \texttt{smlad} instruction: for two 32-bit operands~($a,b$), it computes two signed 16-bit products corresponding to the top~($a_1b_1$) and bottom~($a_0b_0$) parts, respectively, and accumulates their sum into a 32-bit accumulator, $s=a_0b_0 + a_1b_1 + c$. \smallskip

\noindent \textbf{\ourmethod applied to Convolution Feature Map}. The fault is injected at the C implementation level, as it provides sufficient control to activate the backdoor effectively. 

For fault characterization, we first instrument the C implementation to introduce an external GPIO signal indicating the start and end of the \texttt{arm\_convolve\_HWC\_q7\_fast} function. This signal is commonly referred to as the trigger signal. We also temporarily add a UART communication to output the resulting feature map. 

We first perform a normal inference run over a set of input images and record the corresponding feature maps produced by the second layer. These measurements provide the reference feature-map values against which the FI effects can be identified. The UART instrumentation is then removed so that it does not affect the subsequent experiments. 

For selecting effective FI parameters, we scan the window where the trigger signal is high and inject faults with varying glitch strength. To have an effective and repeatable backdoor trigger, whilst remaining stealthy, we aim to fault the smallest number of bytes while having a significant value difference. We minimize the number of faulted bytes to improve stealthiness in the feature map, while aiming for a large value difference to ensure benign byte values do not trigger the backdoor. 

Post characterization, we choose the glitch strength, and fault offset~(time difference between the glitch and the trigger signal) that provides reliable faults. We inject the model backdoor on the specific faulted byte, with a backdoor threshold that will be triggered by this faulted value.

\subsection{\ourmethod Evaluation Metrics} 
\noindent To evaluate the effectiveness of \ourmethod, we use two metrics: the Fault Success Rate (FSR) and the Backdoor Attack Success Rate (ASR). 
Since FI does not always induce the intended fault, we use FSR to measure the probability of successfully achieving the targeted fault outcome. FSR is defined as the percentage of FI trials that successfully achieve the targeted fault outcome, i.e. $({\# \textsf{\small of successful faults}}/{\# \textsf{\small of total fault trials}}) \cdot 100\%$. Here, the total number of trials includes all FI attempts, including resets, unaffected executions, and executions in which the injected fault does not produce the desired outcome.

Our attack subsequently exploits the induced fault to activate a backdoor. Therefore, given a successful fault that produces the intended value change, we measure the Backdoor Attack Success Rate (ASR), as the proportion of such faulted executions in which the backdoor is successfully triggered and produces the intended label.
Note that the faulted input values that lead to targeted misclassification without triggering the backdoor are not counted. We report FSR and ASR for our experiments.

\section{Experimental Evaluation}
\label{sec:experimental_eval}

\begin{figure}[t!]
    \centering
    \includegraphics[width=0.8\linewidth]{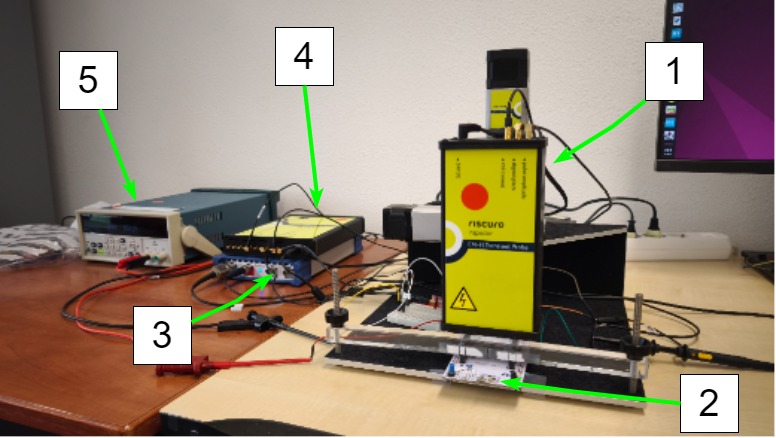}
    \caption{EMFI Setup comprising of: (1) EMFI Unidirectional Transient Probe mounted using the XYZ station; (2) Nucleo-144 target evaluation board; (3) PicoScope; (4) VC Glitcher; (5) External Power Supply.}
    \label{fig:emfiSetup}
\end{figure}

\subsection{Experimental Setup}
\noindent \textbf{Implementation details.}
 \noindent We perform our experiments on embedded 8-bit quantized CNNs implemented using the open source \textit{Neural Network on Microcontroller}~(NNoM) library\footnote{\url{https://github.com/majianjia/nnom/}} combined with the \textit{CMSIS-NN}\footnote{\url{https://github.com/ARM-software/CMSIS-NN}} kernels, which improve NN performance, reflecting real-world conditions.
The implementation is written in C and compiled using {\tt arm-none-eabi-gcc} (version 13.3), with the {\tt -O3} flag enabled to maximize execution speed. \smallskip

\noindent{\bf EMFI Setup.}
We use the Nucleo-144 STM32L4R5ZI-P evaluation board\footnote{\url{https://www.st.com/en/evaluation-tools/nucleo-l4r5zi-p.html}} with an ARM-Cortex-M4 microcontroller~(MCU), a widely used platform for embedded applications. The target is operating at 32 MHz, powered via an external 3.3 V supply and it communicates with a host machine over UART. 
Firmware development and flashing were performed using STM32CubeIDE. 
Evaluation boards such as the one used in our experiments do include capacitors to stabilize the voltage supply to the chip. 
However, this is not an issue for EMFI, since the electromagnetic pulses are generated externally. As such we do not remove any capacitors or modify the development board in any way. 

We use the Keysight Unidirectional FI probe\footnote{\url{https://www.keysight.com/us/en/product/DS1120A/unidirectional-fault-injection-probe.html}} to generate short EM pulses~(see Figure~\ref{fig:emfiSetup}). The probe is mounted on the Precision XYZ Stage, positioning the probe in close proximity to the chip surface without physical contact. 
The probe is driven by a Keysight VC Glitcher DS1160A\footnote{\url{https://www.keysight.com/be/en/product/DS1160A/smartcard-voltage-and-clock-glitcher.html}}, which generates controlled voltage transients synchronized with the target execution, which are transmitted to the EM probe. 
The target board uses General Purpose Input Output~(GPIO) pins to send the trigger signal to the VC Glitcher, indicating the start of the operation to be targeted. 
The trigger output GPIO is configured 

to produce a sharp rising edge to ensure optimal timing precision. 
The VC Glitcher uses this trigger to time the injected glitch. Additionally, the VC Glitcher is connected to a MOSFET via its reset line to fully power-cycle the target when it becomes unresponsive. 
Glitch parameters and target selection are configured using the Inspector software (v2025.12). 
The setup includes a PicoScope oscilloscope to monitor the trigger, coil current or reset lines when needed. 
The coil current informs the effective glitch in the power line applied to the board. 
We use several glitch source power (GSP), where a GSP of 60\% results in voltage dip of -1.7V. \smallskip

\begin{figure}[t]
    \centering    
    \includegraphics[width=0.7\linewidth, trim= 20 40 0 60, clip]{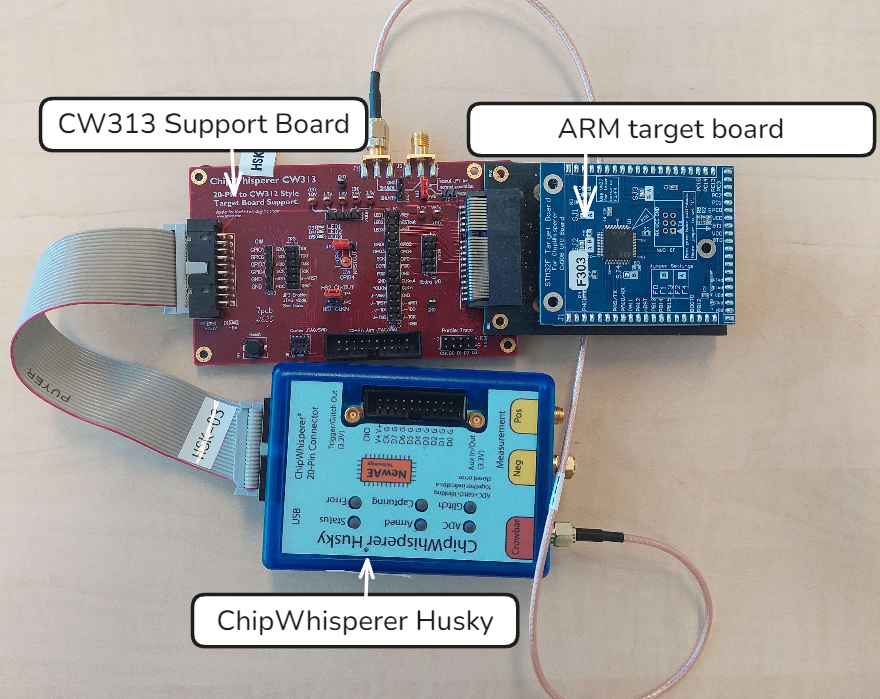}
    \caption{Voltage Glitching experimental setup using ChipWhisperer Husky, CW313 support board and ARM Target.}
    \label{fig:voltage_cw312}
\end{figure}

\begin{figure}
    \centering    \includegraphics[width=1\linewidth]{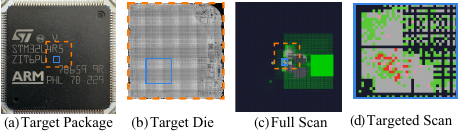}
    \caption{Spatial characterization results for the STM32L4R5ZI MCU for EMFI. (a) packaged chip; (b) actual die position after decapping~(used only to confirm glitches target the chip's die); (c) full scan coalesces multiple coarse and targeted scans with smaller step sizes. (d) targeted scan in the blue boxed region, with the positions marked based on outcomes. Green indicates no faults, grey indicates chip-resets while red indicates successful faults. We choose one of the red points as the final probe position for our experiments.} 
    \label{fig:scan}
\end{figure}

\noindent{\bf Voltage Glitching Setup.}
\noindent For Voltage Glitching experiments, we use NewAE’s ChipWhisperer CW313 board\footnote{\url{https://github.com/newaetech/chipwhisperer-target-cw313}}, which hosts an ARM Cortex-M4 microcontroller clocked at 7.37 MHz for our experiments. Glitches are generated using the ChipWhisperer Husky platform\footnote{\url{https://github.com/newaetech/chipwhisperer-husky}} and injected into the target via the crowbar glitching mechanism, which momentarily shorts the supply voltage to ground. 

Communication between the target and CW-husky takes place over the serial connector, which relays the trigger and reset signals over dedicated pins and also enables serial communication. The setup is depicted in Figure~\ref{fig:voltage_cw312}. \smallskip

\noindent{\bf Software Setup.}
Most off-the shelf setups such as those provided by Keysight and NewAE use UART handler functions to communicate between the host machine and the system under test. 
This technique is also replicated in many custom setups such as described in Bhasin et al.~\cite{cryptoeprint:2025/192, 10.1007/978-3-032-16342-4_3}. 
A drawback of this approach is that calls need to be inserted to these handler functions at each different time point you want this communication to happen, furthermore low-level operations such as reading out the full register state requires additional workarounds. 
Our solution to this is to leverage the on-board debugger and PyOCD to halt the processor at locations of interest which enables arbitrary reads and writes to both registers and memory for debugging purposes. 
While the final attacks do not require this level of control over the processor, it enables us to quickly inspect the processor state at different stages in the calculation to examine the actual fault effects on a deeper level while making fewer modifications to the target code. 

\smallskip

\noindent{\bf Locating the EMFI attacking point.} Figure~\ref{fig:scan} shows the EMFI scan of the target MCU chip, where red positions indicate desired faults during glitching. 
In most cases, a glitch power of 60\% was sufficient to find sensitive regions. It is high enough to induce occasional faults, yet low enough to limit resets (crashes). 
Limiting resets matters because they can mask sensitive regions, especially when higher glitch strengths are used, that a weaker glitch could have exposed otherwise. Given a sensitive region, multiple positions are tested over many measurements to locate a specific position that can reliably produce faults, which in turn is used for the subsequent experiments. 

\begin{table}[!t]
\centering
\caption{MNIST NNoM backdoors: original (clean) accuracy vs.\ backdoor attack success rate (ASR), grouped by attack family. Target class~7.}
\label{tab:mnist-backdoor}
\footnotesize
\setlength{\tabcolsep}{4pt}
\begin{tabular}{lcc}
\toprule
Model & Clean (\%) & ASR (\%) \\
      & (full test set) & ($n{=}100$)\\

\midrule
\multicolumn{3}{@{}l}{\emph{Poison-based}}\\
\quad Pixel & 99.3 & 99.7  \\
\quad FM conv\_1     & 99.4 & 100.0 \\
\quad FM conv\_2    & 99.4 & 100.0 \\
\addlinespace
\multicolumn{3}{@{}l}{\emph{Data-free (weight surgery)}}\\
\quad Data-free FM (conv\_2)  & 97.4 & 99.2 \\
\bottomrule
\end{tabular}
\end{table}

\subsection{Supply-chain Backdoor Injection}
\label{sec:Supply-chain Backdoor Injection}

\noindent\textbf{Poison-based injection.}
The poison-based variant plants the backdoor during training by data
poisoning. We pick a target class $t{=}7$ and a poison ratio $\rho{=}10\%$.
For each poisoned sample we perform the trigger injection \emph{inside} the
forward pass: at the layer that the on-device fault will perturb (e.g.\ the
\texttt{conv2} output in the 5-conv CIFAR-10 model), we overwrite a single
feature map element with the trigger value, and relabel the sample to $t$.
The network is then trained end-to-end on the mixed clean/poisoned set by
minimizing the standard negative log-likelihood, so it learns to associate that
single intermediate activation with class~$t$ while preserving benign accuracy.
After training, the model is quantized and exported to INT8 NNoM weights.
The trigger is chosen so that it maps exactly to $\mathtt{0x7F}{=}127$, making the deployed trigger a single-byte fault in the intermediate layer.

\noindent\textbf{Data-free injection.}
The data-free variant implants the same behavior without any training data or retraining, operating directly on the quantized weights. 
A trigger byte is fixed in the second convolution (\texttt{conv2}) output buffer at pre-defined flat offset. 
We then rewire three layers by weight surgery: in the third
convolution (\texttt{conv3}), for $N_{\mathrm{BD}}{=}16$ output channels we zero all kernel weights and set only the center tap from the trigger channel to $\gamma{=}127$.
In the dense layer we zero one neuron's inputs and connect the
corresponding flatten positions with weight $127$.
In the output layer we set the
target-class row to $+127$ and the others to $-127$. With per-layer right-shift
$9$, injecting $\mathtt{0x7F}{=}127$ at the trigger byte propagates a large
activation along this hand-crafted integer pathway to the target logit, forcing class~$t$, whereas a clean (zero) trigger byte leaves the prediction essentially unchanged. 
No gradient steps are taken, and the benign accuracy of the backdoored model is approximately equal to the clean accuracy of the original model.

\noindent \textbf{Backdoor performance.} Tables~\ref{tab:mnist-backdoor} and~\ref{tab:cifar-backdoor} report the benign accuracy and attack success rate (ASR) of every backdoor variant on MNIST and CIFAR-10. 
ASR reaches $99.2$--$100\%$ on MNIST and $95.3$--$100\%$ on CIFAR-10, while the benign accuracy stays equal to the clean accuracy of each network.

The feature map (FM) FI triggers consistently attain $100\%$ ASR for the non-data-free cases, and $99.2\%$ when Data-free injection is used.
The two attack families perform similarly.
No data-free attack is implemented for CIFAR-10, so all CIFAR-10 entries (Table~\ref{tab:cifar-backdoor}) are poison-based.

\begin{table}[!t]
\centering
\caption{CIFAR-10 NNoM backdoors: original (clean) accuracy vs.\ backdoor attack success rate (ASR). All variants are poison-based. Target class~7.}
\label{tab:cifar-backdoor}
\footnotesize
\setlength{\tabcolsep}{4pt}
\begin{tabular}{llcc}
\toprule
Model & Arch. & Clean (\%) & ASR (\%) \\
& & (full test set) & ($n{=}100$)\\
\midrule
\multicolumn{4}{@{}l}{\emph{Poison-based}}\\
\quad Pixel           & 3conv & 80.0 & 95.3  \\

\quad FM conv\_1  & 3conv & 80.9 & 100.0 \\

\quad FM conv\_1       & 5conv & 83.7 & 100.0 \\
\quad FM conv\_2       & 5conv & 84.4 & 100.0 \\
\bottomrule
\end{tabular}
\end{table}

\begin{figure}
    \centering
    \includegraphics[width=0.75\linewidth]{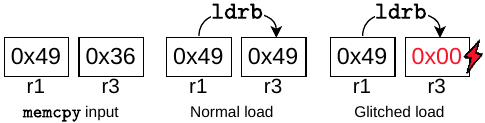}
    \caption{Example of normal and glitched load in \texttt{memcpy}.}
    \label{fig:54_1}
\end{figure}

\subsection{FI Attacks on \texttt{memcpy}}
\subsubsection{FI Characterization on \texttt{memcpy}} 

\noindent \textbf{EMFI on \texttt{load} or \texttt{store} for register data modification.}
We first glitch the load instruction. 
Figure~\ref{fig:54_1} illustrates the data changes. 
In normal execution, the target register \texttt{r3} loads the pixel byte from the address in \texttt{r1}, for example \texttt{0x49}. 
Before this load, \texttt{r3} still holds the previous pixel value, for example \texttt{0x36}. 
Under the glitch, the value in \texttt{r3} becomes \texttt{0x00}. 
It indicates that the load is not skipped and its control flow is not disrupted, because a skipped load would leave the residual value \texttt{0x36} in \texttt{r3}, whereas the observed \texttt{0x00} is a direct corruption of the register content. 
The same behavior is observed when glitching the \texttt{strb} instruction, where the stored pixel value likewise becomes \texttt{0x00}.
Similar glitching effects have also been observed on cryptographic implementations~\cite{menu:hal-02338456, Moro_2013}.
Our characterization shows that this fault-induced data change is predictable, so adversaries can exploit it.

\begin{figure}[!t]
    \centering
    \includegraphics[width=1\linewidth]{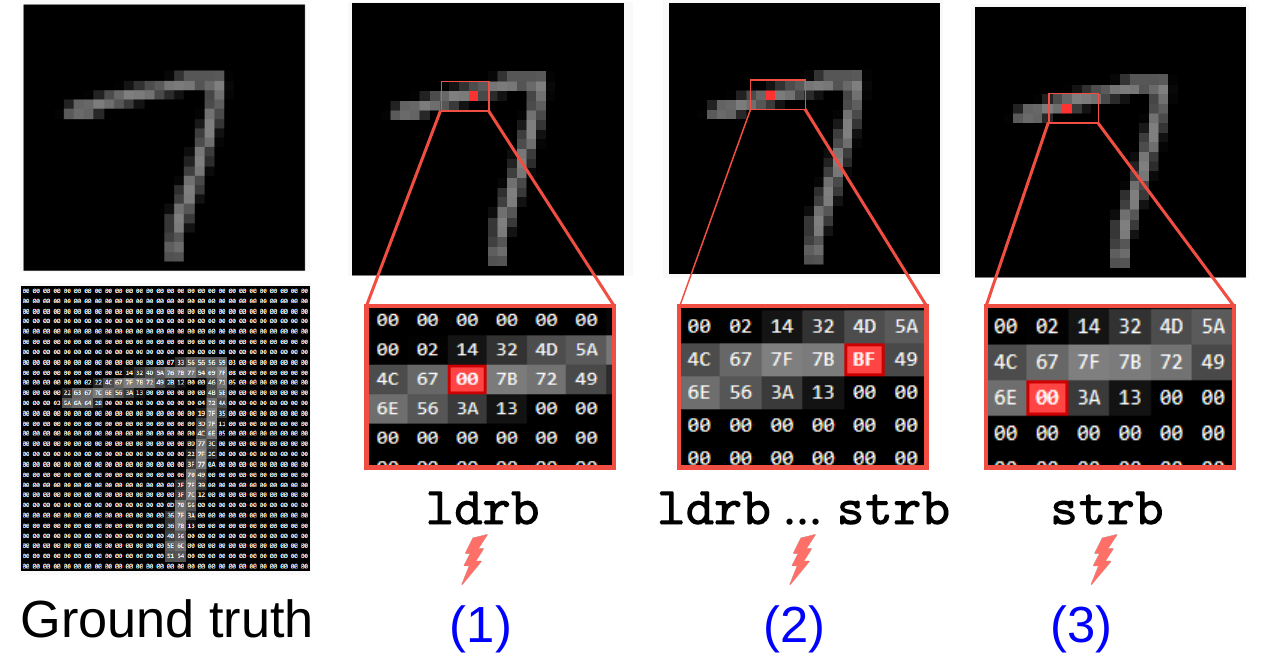}
    \caption{EMFI outcomes targeting the \texttt{memcpy} operation on an MNIST image of class label 7, along with the ground truth (true MNIST image and a grid overlay showing all pixel byte values). Glitching \textsf{(1)} \texttt{ldrb} or \textsf{(3)} \texttt{strb} causes the target byte to be zeroed; Glitching \textsf{(2)}, between the \texttt{ldrb} and \texttt{strb}, results in pixel value \texttt{0xBF} as a result of glitching the  operation.}
    \label{fig:emfiResultsMnistVisua0l}
\end{figure}

\noindent \textbf{EMFI between \texttt{load} and \texttt{store} for register data modification.} 
We also glitch between the \texttt{ldrb} and \texttt{strb} instructions. 
We observe that when we put \texttt{NOPs} between \texttt{ldrb} and \texttt{strb}, the target register holds a faulted value \texttt{0xBF} (see Figure~\ref{fig:emfiResultsMnistVisua0l}), which is the high byte encoding of a NOP after a glitch.
When we replace 1 NOP with \texttt{orr r7, r7, r7} or \texttt{add r7, r7, \#0}, we observe \texttt{0x1d} or \texttt{0x64} in most cases. 
Experimental results demonstrate that the target register either gets the address stored in the destination register or the source register. 
As such, we get a predictable fault model if the starting memory address is known, and backdoor attacks could exploit it. 
We include technical details of the analysis in Appendix~\ref{appendix:memcpy}.

\noindent \textbf{Voltage glitching targeting the \texttt{memcpy}.} 
We perform a characterization for voltage glitching on the C implementation of the \texttt{memcpy} function. Table~\ref{tab:single_byte_fault_clusters} lists the identified fault parameters corresponding to stable byte changes. Note that FI characterization in the voltage glitching setup is less precise than in the EMFI setup, but it can still be exploited by an adversary to trigger a backdoor.

\begin{table}[t]
\centering
\caption{Voltage glitching parameters and outcomes for single-byte fault effects observed across multiple FI trials.}
\label{tab:single_byte_fault_clusters}
\begin{tabular}{lccc}
\toprule
Offset pattern & Trials & Changed index & Original $\rightarrow$ New \\
\midrule
262--267& 15 & 201 & $0 \rightarrow 87$ \\
317--327 & 12 & 200 & $0 \rightarrow 71/79/95$ \\
302 & 10 & 200 & $0 \rightarrow 127$ \\
\bottomrule
\end{tabular}
\end{table}

\subsubsection{FI-triggered backdoor attacks targeting \texttt{memcpy}} 
Following the FI characterization, we inject the backdoor in
the 201st pixel of the input image.
The backdoor is triggered when that pixel value exceeds 83, where the faulted value serves as the backdoor trigger. 
In particular, the EMFI-based trigger targets the fault induced between the \texttt{ldrb} and \texttt{strb} instructions, whereas the voltage glitch-based trigger targets a fault applied directly to the
\texttt{memcpy} that copies the 201st pixel, to gain their respective, sufficient level of control. 
It can be observed in Table~\ref{tab:pixelbd} that the two configurations reach FSR of 34.3\% (EMFI) and 32.2\% (voltage glitch).
In both cases the attack success rate (ASR) is 100\%, indicating that once the fault triggers the backdoor, all test images are classified into the target class.
Note that the faults do not affect the behavior of non-backdoored models, as the induced data changes are confined to a single byte.

\begin{table}[!t]
\centering
\caption{Attack success rates for pixel backdoor on MNIST for FI~(for target class 7). Accuracy refers to the benign accuracy on the \emph{full test set}. The backdoor attack success rate (ASR) and fault success rate~(FSR) are measured on \emph{100 test cases}.}
\label{tab:pixelbd}
\footnotesize
\setlength{\tabcolsep}{4pt}
\begin{tabular}{llccc}
\toprule
Dataset & FI technique & Accuracy (\%) & FSR (\%) & ASR (\%) \\
        &              & (full test set) & ($n{=}100$) & ($n{=}100$) \\
\midrule
\multirow{2}{*}{MNIST}    & EMFI              & \multirow{2}{*}{99.30} & 34.3& 100.0 \\
                          & Voltage Glitching &                        & 32.2 & 100.0\\

\bottomrule
\end{tabular}
\end{table}

\subsubsection{Additional analysis targeting \texttt{memcpy}}
In the case of adding \texttt{NOPs} between \texttt{ldrb} and \texttt{strb}, faulting in the NOP sled results in the faulted pixel value being \texttt{0xBF}. 
This result could be explained by the ARM 16-bit Thumb instruction: the NOP encoding is \texttt{0xBF00}\footnote{See page 306 of the Armv7-M Architecture Reference Manual: \url{https://developer.arm.com/documentation/ddi0403/ed/}}. 
Hence, the faulted pixel's value is being overwritten with the high byte of the NOP encoding. 
During the trigger window, only this NOP sled is being executed. 
The Data bus must therefore be inactive, while the Instruction bus is active, it is fetching NOP instructions. 
We conjecture that in this case, the EMFI-induced glitch resulted in leakage of this \texttt{NOP} opcode.

\begin{figure}[t]
    \centering

        \includegraphics[width=0.75\linewidth, trim= 0 40 0 10, clip]{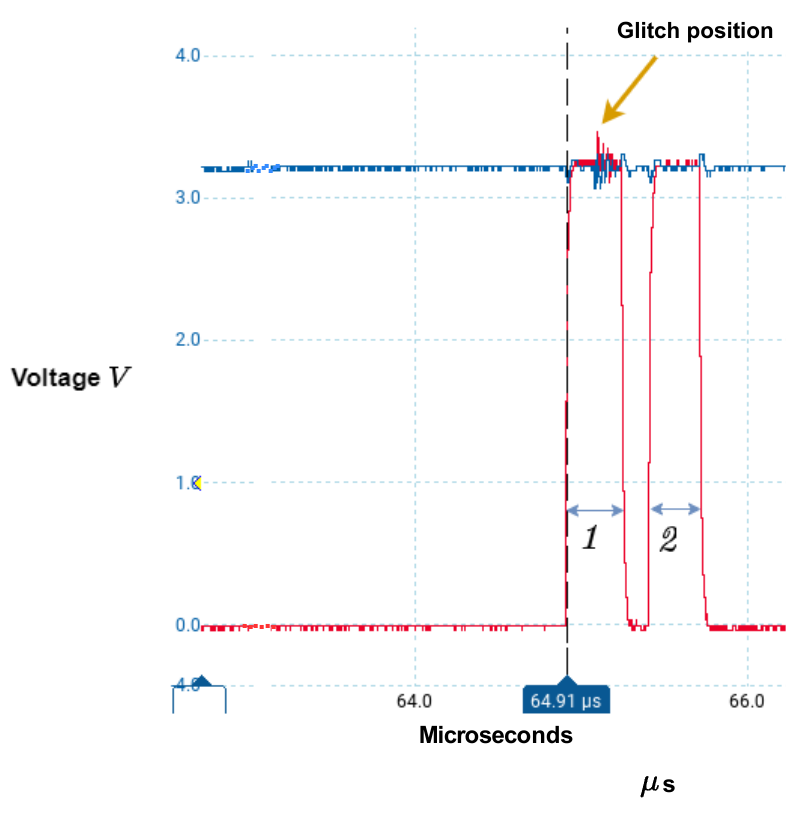}

        \caption{Power trace targeting the \texttt{smlad} operation in convolutional layer 2. The blue line is the main trigger. The red line indicates the pre-trigger is used to locate \texttt{smlad}. The two pulses correspond to two \texttt{smlad} code blocks~(1 and 2) in Figure~\ref{fig:mapping}.} 
        \label{fig:smlad-picoscope}
 
\end{figure}

\begin{figure}[t]
  \centering
  \resizebox{\columnwidth}{!}{
  \begin{tikzpicture}[
    font=\ttfamily\small,
    line/.style   = {anchor=west},
    cmt/.style    = {anchor=west, text=black!50},
    dots/.style   = {anchor=west, text=black!55},
    target/.style = {draw=red!70!black, line width=0.8pt, rounded corners=2pt,
                     inner sep=3pt, anchor=west},
  ]

    \begin{scope}[on background layer]
      \fill[blue!6,  rounded corners=4pt] (-0.30, 1.00) rectangle (7.10,-5.5); 
      \fill[green!8, rounded corners=4pt] ( 7.30, 1.00) rectangle (12.90,-5.5); 
    \end{scope}

    \node[anchor=west, font=\sffamily\small\bfseries] at (0,0.6)   {C source};
    \node[anchor=west, font=\sffamily\small\bfseries] at (7.5,0.6) {Disassembly};

    \node[line] at (0, 0.0)             {while (colCnt) \{};
    \node[cmt]  at (0.4,-0.5)           {/* block 1 */};
    \node[cmt]  at (0.4,-1.0)           {// set Pre-trigger High};
    \node[dots] at (0.6,-1.3)           {$\vdots$};
    \node[target] (cT) at (0.4,-2.0)    {sum4 = \_\_SMLAD(inA21, inB2, sum4);};
    \node[cmt]  at (0.4,-2.5)           {// set Pre-trigger low};
    \node[dots] at (0.6,-3.0)           {$\vdots$};
    \node[cmt]  at (0.4,-3.5)           {/* block 2 */};
    \node[dots] at (0.6,-4.0)           {$\dots$};
    \node[line] at (0,-4.5)             {\}};

    \node[dots] at (7.7,0)              {$\vdots$};
    \node[line] at (7.5,-0.7)           {smlad~~r3, r1, ip, r3};
    \node[line] at (7.5,-1.1)           {smlad~~r8, r2, sl, r8};
    \node[line] at (7.5,-1.5)           {ldr~~~~r1, [sp, \#4]\quad\quad\textcolor{red}{(a)}};
    \node[target] (aT) at (7.5,-2)      {smlad~~ip, r2, ip, r1\quad\textcolor{red}{(b)}};
    \node[line] at (7.5,-2.5)           {str.w~~fp, [r6, \#40]\quad\;\;\textcolor{red}{(c)}};
    \node[dots] at (7.7,-2.9)           {$\vdots$};

    \draw[-{Stealth[length=2.5mm]}, line width=0.8pt]
      (cT.south) to[out=-15,in=221,looseness=0.5] (aT.west);

    \node[anchor=north, fill=white, rounded corners=2pt, inner sep=2pt] at (9.3,-3.4) {
      \normalfont
      \begin{tabular}{cl}
        \toprule
        \multicolumn{2}{c}{\sffamily\footnotesize Resulting value after skipping different instructions}\\
        \midrule
        \textcolor{red}{(a)} & \texttt{0x80}~~($-128$)\\
        \textcolor{red}{(b)} & \texttt{\textcolor{red}{0x7F}}~~($+127$)\\
        \textcolor{red}{(c)} & \texttt{0xFF}~~($-1$)\\
        \bottomrule
      \end{tabular}
    };
  \end{tikzpicture}}
  \caption{Mapping the CMSIS-NN \texttt{smlad} C operations to its Cortex-M4 disassembly version, optimized with GCC {\tt -O3}. The two \texttt{smlad} groups correspond to blocks~1 and~2 in the power
  trace (Figure~\ref{fig:smlad-picoscope}). Faulting the boxed \texttt{smlad} at \textcolor{red}{(b)} results in the empirically observed value \texttt{0x7F} (=127) in the feature map, whereas targeting the instruction before and after the \texttt{smlad},  \textcolor{red}{(a)} and \textcolor{red}{(c)}, yields different results; confirming the EMFI attack targeted the \texttt{smlad} instruction.}
  \label{fig:mapping}
\end{figure}

\subsection{FI Attacks on Convolutional Layers}
\label{sec:fi_conv}

\noindent In this section, we conduct fault injection attacks on convolutional layers with a focus on the convolution operation and especially the \texttt{SMLAD} instruction used in CMSIS-NN kernels that is specifically designed for edge neural network acceleration. 
We first glitch the \texttt{SMLAD} instruction, since the target implementation uses several \texttt{SMLAD} blocks to optimize the inference speed.
We also glitch higher-level convolution operations, focusing on the C implementation.

\subsubsection{FI Characterization on \texttt{SMLAD}}

\noindent\textbf{Locating the \texttt{SMLAD} instruction.} SMLAD resides at the core of the CMSIS-NN kernels. 
In particular, we target \texttt{arm\_nn\_mat\_mult\_kernel\_q7\_q15\_reordered} that consists of two blocks of \texttt{SMLAD} executions, both containing 4 consecutive \texttt{SMLAD} calls. 

Our fault is set to be triggered during the first execution of the \texttt{SMLAD} block.

Figure~\ref{fig:mapping} shows the first \texttt{SMLAD} execution, where \texttt{sum4} indicates the fourth \texttt{SMLAD} call whose corresponding disassembly is also shown on the right. 
To further analyze the glitch pattern, we add pre-triggers around the \texttt{SMLAD} code blocks, which allow us to pinpoint the exact \texttt{SMLAD} call being targeted when adjusting the glitch offset.

\noindent \textbf{EMFI Glitching.} 
Different glitch outputs are observed when we adjust the glitch parameters. 
Once we observe faults in the registers, we take a smaller glitch-parameter step to inspect them further. 
In the EMFI, various glitched patterns are observed, including massive corruption, small-scale data changes, and byte-level data changes. 
After the scanning, we select one repeatable one-byte data change to high value as the fault-induced data change to exploit. 
The trace in Figure~\ref{fig:smlad-picoscope} confirms that the glitch point is inside the first \texttt{SMLAD} execution, and timing suggests that we are targeting the fourth \texttt{SMLAD} call in the first block as shown in Figure~\ref{fig:mapping}.
It can be observed that faulting boxed \texttt{SMLAD} results in \texttt{0x7F}.

\begin{figure}
    \centering
    \includegraphics[width=0.9\linewidth, trim=0 156 0 0, clip]{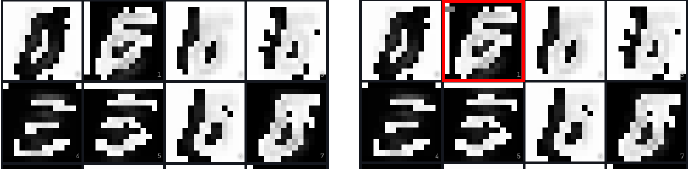}
    \caption{Output feature maps of the second convolution (the conv2 backdoor's trigger layer) of the MNIST model for one input: clean (left) vs.\ the backdoor activated on the device (right). Each tile is one of the output channels ($8/24$ depicted). Fault alters one top-left byte in a single channel (red square), while the other channels remain unaltered.}
    \label{fig:FM_FI}
\end{figure}

\begin{figure}
    \centering
    \includegraphics[width=1\linewidth]{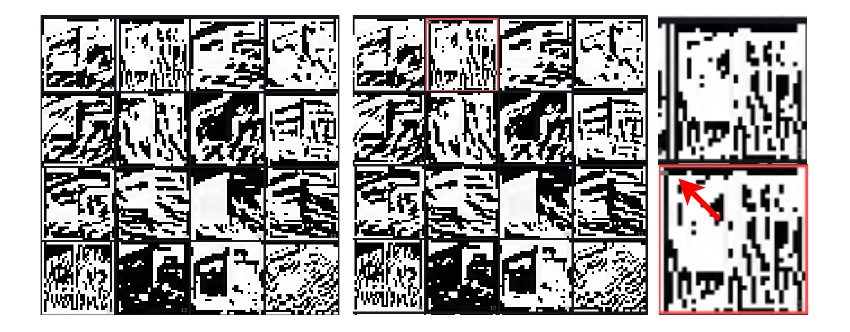}
    \caption{Feature map view of the \texttt{conv2} output of the CIFAR-10 5-conv model under the feature map fault injection backdoor (target class~7). \textbf{Left:} the channels of \texttt{conv2} for a clean input (trigger inactive). \textbf{Middle:} the same channels when the backdoor is activated. Only a single channel (red box) is altered while every other channel is bit-identical to the clean case. \textbf{Right:} magnified channel~1, clean (top) vs.\ activated (bottom). The grey pixels (indicated by the red arrow) mark the one injected data change. }\label{fig:placeholder}
\end{figure}

Figure~\ref{fig:FM_FI} shows $8$ out of $24$ output channels of the second convolution for the same image, under normal inference (left) and with the trigger applied on the device (right). 
Visually the two activation grids are indistinguishable, since they are identical in every channel except one. 
Only channel~1 (highlighted in red) differs, and within that channel only a single spatial location is overwritten (see Figure~\ref{fig:placeholder} for the same effect on CIFAR-10).
This is the precise one-byte fault into an intermediate feature map, after which inference continues normally and the network is driven to the target class. Since the perturbation is confined to a single intermediate value with the input image untouched, it remains undetected by input-space backdoor defenses.

\begin{figure}[tb]
    \centering

    \begin{subfigure}[b]{0.85\linewidth}
        \centering
        \includegraphics[width=0.65\linewidth, trim=0 5 0 15, clip]{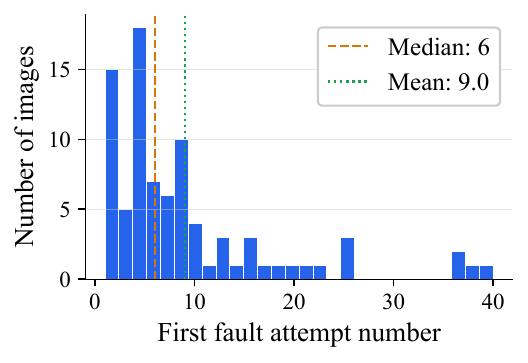}
        \caption{Distribution of first-successful-fault attempts. 76\% of first successful faults occurred within 10 attempts.} 
        \label{fig:minimal_FIs_hist}
    \end{subfigure}

    \hfill

    \begin{subfigure}[b]{0.8\linewidth}
        \includegraphics[width=\linewidth]{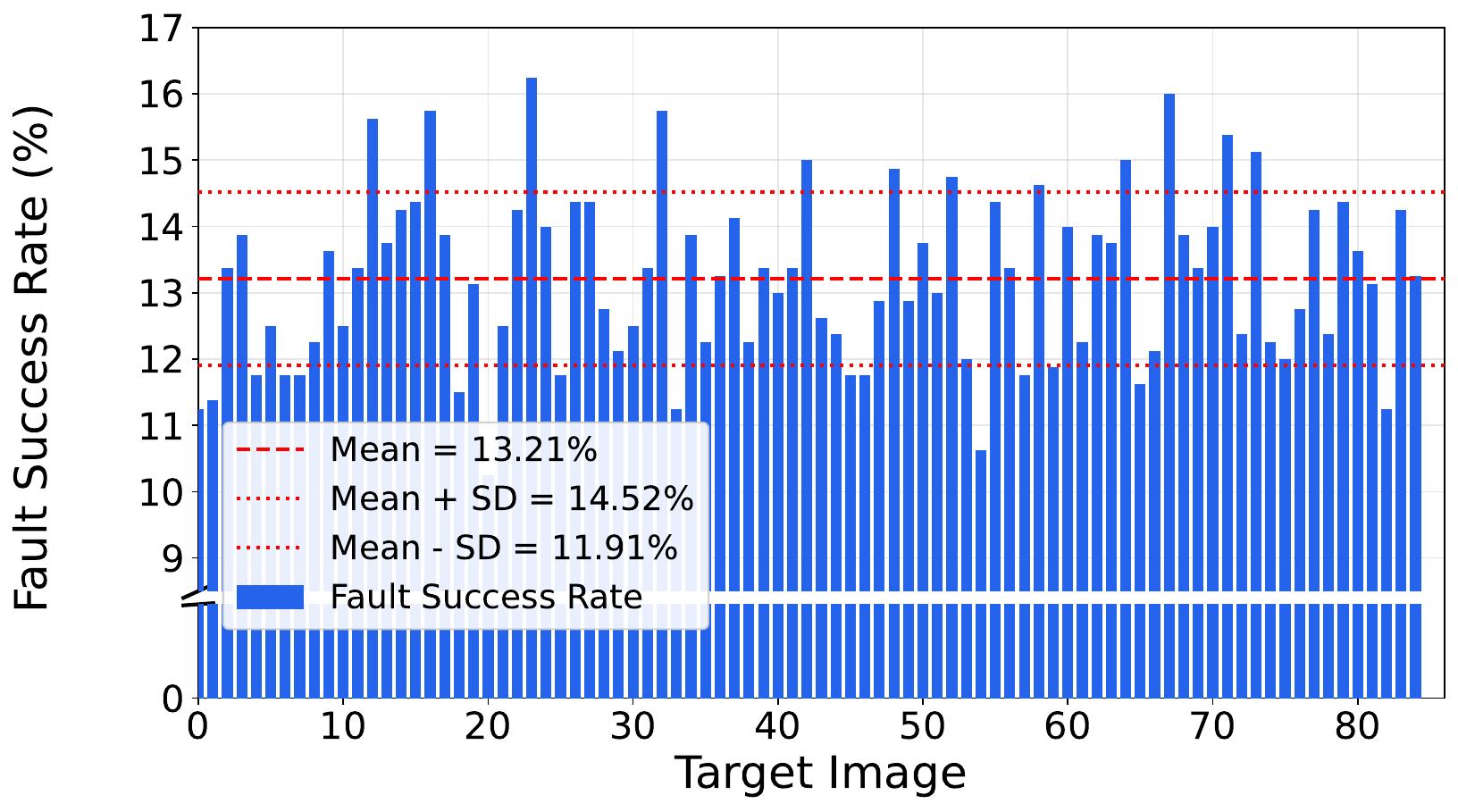}
        \caption{FSR distribution per image.}
        \label{fig:second_figure}
    \end{subfigure}

    \caption{FSR results for 100 MNIST images (over 800 measurements each) targeting \texttt{SMLAD} via EMFI, excluding images whose correct class matches the intended fault class.}
    \label{fig:combined}
\end{figure}

\subsubsection{FI-triggered backdoor attacks targeting \texttt{SMLAD}}
In Table~\ref{tab:fi_comparison}, the \texttt{SMLAD}\;(\texttt{ASM}) row targets the \texttt {conv\_2} layer that is optimized by the \texttt{SMLAD} instruction, where the backdoor is injected by data-free method. 
It can be observed that the attack is effective, but the FSR (13\%) is relatively low.
Figure~\ref{fig:minimal_FIs_hist} shows that the first successful fault occurred after a median of 6 glitch tries (mean 9).

We also demonstrate that the FSR is stable across different images in Figure~\ref{fig:second_figure}.
Even though the FSR is low, the backdoor can still be triggered through EMFI, and the data-free backdoor remains fully effective on it.

\begin{table}[!t] 
\centering
\caption{Fault injection-triggered backdoors.}
\label{tab:fi_comparison}
\scriptsize
\setlength{\tabcolsep}{2.5pt}
\renewcommand{\arraystretch}{1.05}
\resizebox{\columnwidth}{!}{
\begin{tabular}{lccccc}
\toprule

\multirow{2}{*}{Dataset} &

\multirow{2}{*}{Tar. Layer} &

\multirow{2}{*}{Tar. Impl.} &

\multirow{2}{*}{Backdoor} &

\multicolumn{2}{c}{EMFI / VG} \\

\cmidrule(lr){5-6}

& & & & FSR (\%) & ASR (\%) \\

\midrule
MNIST    & conv\_1 & \texttt{*\_basic(C)}       & Poison      & 20.2/99.9 & 100.0/100.0 \\
MNIST    & conv\_2 & \texttt{*\_fast(C)}     & Poison+data-free & 17.8/100.0 & 100.0/100.0 \\
MNIST    & conv\_2 & \texttt{SMLAD(ASM)}        & data-free        & 13.0/N/A    & 100.0/N/A \\

CIFAR-10    & conv\_2 &\texttt{*\_fast(C)}  & data-free        & 5.8/100.0 & 100.0/100.0 \\

\bottomrule
\end{tabular}
} 
\label{fig:fsr-asr}
\end{table}

\subsubsection{Additional analysis targeting \texttt{SMLAD}}

Due to the \texttt{-O3} optimization, the disassembled version of \texttt{arm\_nn\_mat\_mult\_kernel\_q7\_q15\_reordered} is complex. 
The \texttt{SMLAD} part is recognizable, though not a one-to-one mapping to the C code (see Figure~\ref{fig:mapping}). 
Because of the pre-trigger placement (see Figure~\ref{fig:smlad-picoscope}), where we can see the two blocks of 4 \texttt{SMLADs} being called, we know that we are roughly glitching during \texttt{SMLAD} execution. 
We use simulation to further explore the root cause of the physical fault injection. For this, we feed the software state of our target device to an instrumented ARM emulator to produce execution results for any skipped instruction.
Simulated faults are rudimentary in that it simply skips faulted instructions and checks the resulting difference in the output layer, discarding any faults that result in a crash/reset. 
Besides considering instruction skips, the simulator also reports data corruptions~(by tracing the register contents) that may or may not influence the program execution.
Using this simulation, we found that when glitching the fourth \texttt{smlad}, we get the expected \texttt{0x7F} from our physical attack experiments, while glitching the instructions immediately before and after the \texttt{smlad} gives different results, supporting the hypothesis that we are indeed glitching the instruction we are targeting.

\subsubsection{FI-triggered backdoor attacks targeting \texttt{arm\_convolve\_HWC\_q7\_basic}}

We also target higher-level convolution operations, in the C implementation.
We characterize the byte at position 0 of the {\tt conv\_1} output feature map as the target byte, as successful faults consistently change its value to \texttt{0x7F}. 
To exploit, a backdoored model was evaluated on an MNIST dataset. We obtain 20.2\% FSR and 100.0\% ASR, reported in Table~\ref{tab:fi_comparison}.
We also observe that a single \ourmethod-style physical fault does not influence the performance of a normal model. 
The model performs normally with the original accuracy, as NN are robust to small noise.

\subsection{Evaluation against Backdoor Detection}

\noindent Table~\ref{tab:def-mnist-cifar} summarizes the performance of four representative defenses (i.e., Neural Cleanse, STRIP, Activation Clustering, ABS) against our method on MNIST and CIFAR-10. 
In every configuration the attack is both effective and benign-preserving (ASR $\approx100\%$ at benign accuracy near the baseline), so detectability is the discriminating factor, and it falls sharply as the trigger moves from the input into deeper feature maps. 
On MNIST, the pixel backdoor is flagged by $4/4$ defenses, the \texttt{conv\_1} fault by $3/4$, and the \texttt{conv\_2} fault by only $1/4$.
On CIFAR-10, the pixel backdoor is caught by $3/4$, \texttt{conv1} by $1/4$, and \texttt{conv2} by none. 
The reason is that the input-space defenses (STRIP, Neural Cleanse, ABS) degrade or fail outright on feature map triggers, because the input image is never modified and the fault is injected only at run time.
The deepest \texttt{conv\_2} trigger therefore evades all of them. 
These results confirm that feature map fault injection backdoors, especially at deeper layers, are substantially stealthier than conventional pixel backdoors.

\begin{figure}[!t]
    \centering
    \includegraphics[width=0.75\linewidth, trim= 0 15 0 30, clip]{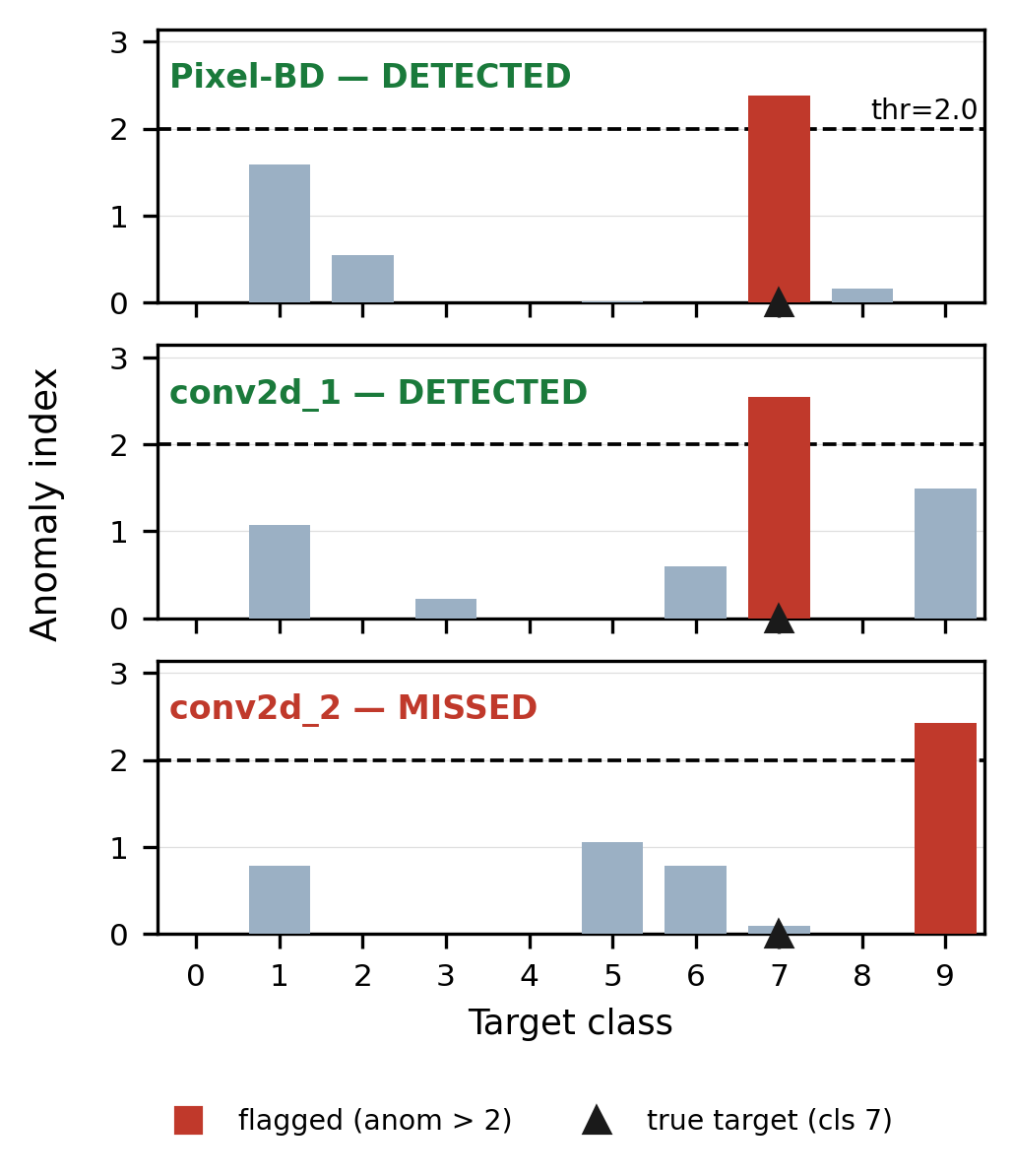}
    \caption{Neural Cleanse anomaly index per candidate target class for the three CIFAR-10 backdoors. A class above the MAD threshold (dashed, $2.0$) is flagged~(red); $\blacktriangle$ is the true target (class~7). Pixel-BD and conv1 are detected ($2.38$, $2.54$); conv2 is missed. 
}

    \label{fig:NC_show}
\end{figure}

\begin{figure}[!t]
    \centering
    \includegraphics[width=1\linewidth]{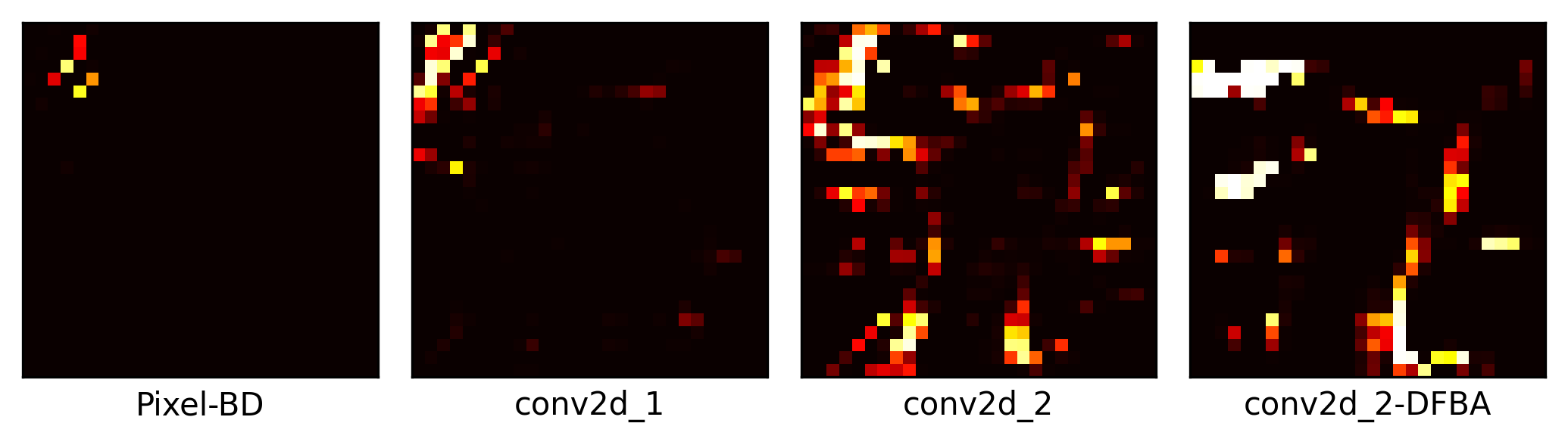}
    \caption{Neural Cleanse recovered trigger (class~7) for the four MNIST models. Only the pixel backdoor yields a localized trigger.}

    \label{fig:ncrec}
\end{figure}

\noindent \textbf{Case analysis on Neural Cleanse.} Figure~\ref{fig:NC_show} provides a case study on Neural Cleanse (NC).
NC reverse-engineers, for each candidate class, the smallest input-space trigger that forces that class, and flags any class whose trigger is anomalously small (MAD index $>2$). 
Figure~\ref{fig:ncrec} shows the recovered triggers by NC, where NC falls short for feature map backdoors.
For the pixel backdoor this works as intended where the recovered trigger is sparse and localized exactly on the planted pixel, so the target class is correctly identified. 
For the feature map backdoors, however, NC fails to detect the deeper \texttt{conv\_2} fault-induced backdoors  and sometimes flags an unrelated class instead (e.g., class~9). 
This case study shows that NC succeeds only when an input-space equivalent of the trigger exists but fails when it goes to a deeper layer.

\begin{table}[!t]
\centering
\caption{Backdoor detection on MNIST and CIFAR-10
NNoM models (target class~7).
\ding{51}: detected; \ding{55}: missed.}
\label{tab:def-mnist-cifar}
\footnotesize
\setlength{\tabcolsep}{2.5pt}
\resizebox{\columnwidth}{!}{
\begin{tabular}{@{}lccc@{\hspace{5pt}}|@{\hspace{5pt}}ccc@{}}
\toprule
& \multicolumn{3}{c}{MNIST}
& \multicolumn{3}{c}{CIFAR-10} \\
\cmidrule(lr){2-4}\cmidrule(lr){5-7}
& Pixel & FM conv\_1 & FM conv\_2 & Pixel & FM conv\_1 & FM conv\_2 \\
\midrule
Benign acc.\ (\%)
& 99.30 & 99.43 & 99.37 & 79.98 & 83.71 & 84.37 \\
ASR (\%)
& 99.70 & 100.0 & 100.0 & 95.30 & 100.0 & 100.0 \\
\midrule
Neural Cleanse
& \ding{51} & \ding{51} & \ding{55}
& \ding{51} & \ding{51} & \ding{55} \\
STRIP
& \ding{51} & \ding{55} & \ding{55}
& \ding{51} & \ding{55} & \ding{55} \\
Activation Clust.
& \ding{51} & \ding{51} & \ding{51}
& \ding{51} & \ding{55} & \ding{55} \\
ABS
& \ding{51} & \ding{51} & \ding{55}
& \ding{55} & \ding{55} & \ding{55} \\
\midrule
Detected
& 4/4 & 3/4 & 1/4 & 3/4 & 1/4 & 0/4 \\
\bottomrule
\end{tabular}
}
\end{table}

\noindent \textbf{Case analysis on adaptive STRIP.} When STRIP is unaware of the feature map trigger, the feature map backdoor cannot be detected but operates normally. 
When STRIP injects the fault during probing, it detects the backdoor almost perfectly, yet the same injection causes most probed inputs to be falsely rejected at approximately $100\%$. 
As a result, adapting the defense to the feature map trigger destroys its regular detection, making it unstable in practice.
However, we believe that an adaptive defense that balances regular backdoor detection, clean-input accuracy, and feature map backdoor detection is possible, and we leave its design to future work.

\ourmethod is stealthy because the trigger resides in an intermediate feature map and fires only under physical fault injection, leaving no input-space artifact.
This stealth deepens when the backdoor is confined to a later layer such as \texttt{conv\_2}, which is harder for defenses to locate, and by targeting the \texttt{SMLAD} instruction, we show the attack generalizes to real, optimized deployments.

\subsection{Generalization of \ourmethod to other platforms}
\noindent \ourmethod is designed to generalize across different implementations by identifying points/locations in the implementation where a single value change can act as a trigger for the backdoored model, rather than relying on specific assembly instructions. 
Regardless of the architecture, implementation or compiler, when a NN model is implemented in software, values must necessarily be loaded, stored and processed, including the accumulation of multiplication results. 
Consequently, operations such as \texttt{ldrb}, \texttt{strb}, and \texttt{smlad} represent potential attack vectors across all software implementations. These operations are supported by a broad range of modern ARM processors, including microcontrollers based on ARMv7-M/ARMv8-M and application processors based on ARMv7-A/ARMv8-A, which are widely used in embedded systems, smartphones, and other computing platforms. Our \textit{characterize-then-exploit} approach allows the adversary to account for target-specific microarchitectural differences during the characterization phase. Our attack can therefore be adapted to the specific target without changing the underlying methodology.

\section{Countermeasures}

\noindent Countermeasures against FI attacks can be deployed at different levels of the implementation stack. Traditional approaches, proposed for cryptographic implementations, include hardware-based fault detection circuits~\cite{baksi2022faultsurvey, BarenghiBKN12} and error correction mechanisms~\cite{ShepherdMHAGHN21, Bar-ElCNTW06}. Another widely used strategy is the introduction of computational redundancy, where critical operations are executed multiple times and their outputs are compared to detect inconsistencies.

\noindent {\bf Randomized Self-Reduction (RSR).} Randomized Self-Reduction (RSR) is a software-level countermeasure based on the principle of random self-reducibility~\cite{DBLP:journals/jcss/BlumLR93}. Instead of executing a sensitive operation directly, the original computation is transformed into multiple randomized instances of the same operation. These instances operate on random inputs or randomized versions of the original input, and their outputs are subsequently combined to recover the correct result. Since an adversary must induce consistent faults across multiple randomized computations, successful FI becomes significantly more difficult.

Several RSR constructions for modular arithmetic operations commonly used in cryptographic implementations have been proposed in~\cite{DBLP:conf/iccad/ErataCENRRPA0S24}, enabling protection of schemes such as RSA-CRT and Kyber key generation. More recently, the concept has been extended to TinyML workloads~\cite{etimfault2026}, where RSR variants have been developed for neural network primitives including the sigmoid activation function, matrix--vector multiplication, matrix--matrix multiplication, and convolution operations. However, the applicability and effectiveness of RSR depend on whether the target operation possesses suitable random self-reducibility properties.

\noindent {\bf Targeted Countermeasures for Neural Network Primitives.} In addition to generic fault-tolerance mechanisms, several countermeasures have been proposed for specific neural network operations. For fault injection attacks targeting the softmax activation function, 
\cite{e105-a_3_300} proposes randomizing the initialization value of the summation accumulator. The implementation further verifies whether the expected number of loop iterations has been executed, enabling detection of control-flow perturbations that skip computations. Another lightweight consistency check exploits the fact that softmax outputs form a probability distribution: the sum of all output probabilities should be close to one. Deviations from this property can therefore be used to detect faults affecting the softmax computation.

\subsection{RSR implementation of \texttt{memcpy}}
\noindent Since \texttt{memcpy} copies bytes and effectively realizes the identity function $f(x)=x$, we use the RSR identity $x=(x+r)-r$ for a random value $r$. We execute two independent randomized instances using fresh random values {\tt rand1} and {\tt rand2}. 
Listing~\ref{lst:rsr_memcpy} illustrates the assembly-level implementation. The first instance loads $x$ from the source address and applies the random value \texttt{rand1}, while the second instance independently loads $x$ and uses \texttt{rand2}. The resulting values are compared using \texttt{cmp}. A mismatch indicates that the two randomized instances have produced different results and consequently signals a fault through the  \texttt{fault\_detected} branch. Since the loads are performed independently, a fault affecting one load can be detected through the inconsistency between the two instances.

The implementation also verifies the final store operation. After the consistency check, the verified value is written to the destination using \texttt{strb}. The stored value is subsequently loaded back using \texttt{ldrb} and compared against the expected value. Hence, a fault that corrupts the value during the store operation is also detected. This post-store verification is particularly important for \texttt{memcpy}, as the fault injection attack can target either the load or the store instruction. Therefore, the complete implementation provides detection for faults affecting both stages of the \texttt{memcpy} operation: \emph{(i)} corruption of the loaded source value and \emph{(ii)} corruption of the value written to the destination. \smallskip

\noindent {\bf Implementation Specifics and Practical Validation.} To identify faults detected by the countermeasure, we introduce a distinct modification in the image by setting all subsequent pixel-byte values to \texttt{0xFF}. This allows us to distinguish countermeasure-detected faults from normal executions and determine whether the injected fault was successfully captured. We perform EMFI on the RSR-protected \texttt{memcpy} for 500 measurements at a fixed characterized fault offset. Of these measurements, 21.4\% triggered the RSR \texttt{fault\_detected} mechanism, while no successful fault is observed with a single fault injection.  \smallskip

\begin{figure}
    \centering
    \begin{minipage}{0.36\linewidth}
        \centering
        \subfloat[at 300\;ns]{\label{fig:ascon_corr1}\includegraphics[width=\linewidth, trim=0 0 0 0, clip]{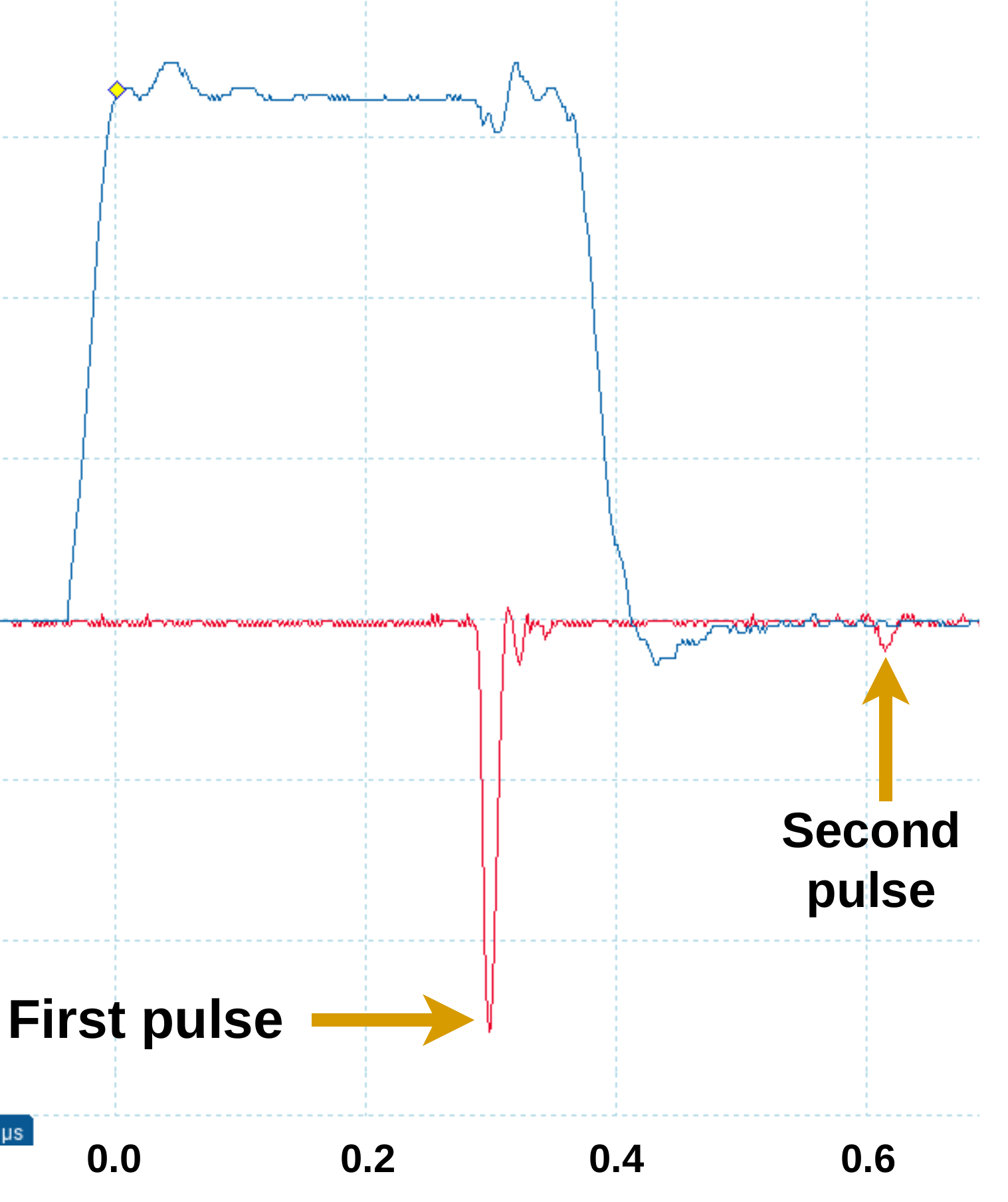}}
    \end{minipage}\hfill
    \begin{minipage}{0.61\linewidth}
        \centering
        \subfloat[at 800\;ns]{\label{fig:ascon_corr2}\includegraphics[width=\linewidth, trim=0 0 0 0, clip]{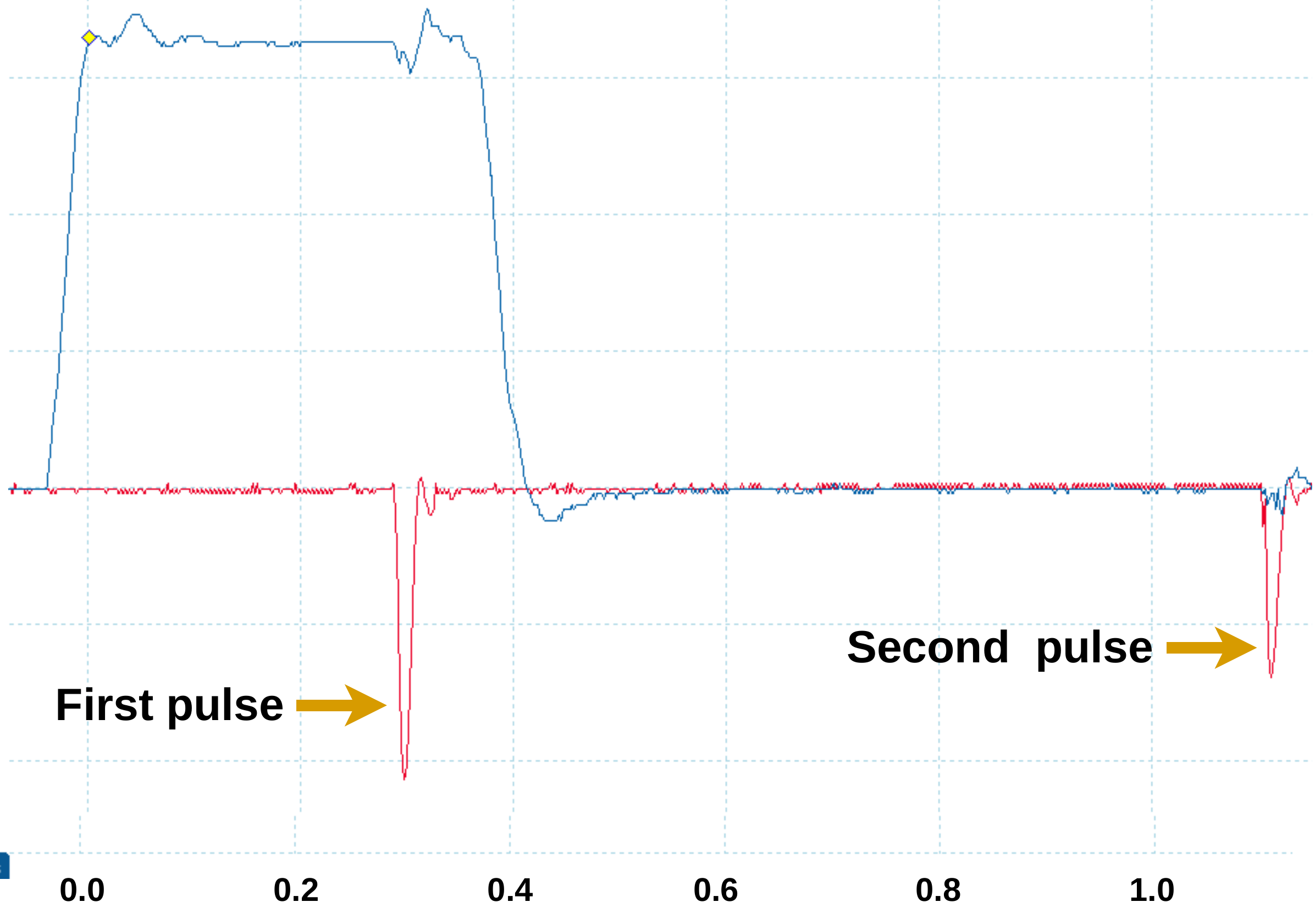}}
    \end{minipage}
    \caption{Time between two consecutive glitches, showing the difference in glitch strength of the second pulse at different offsets.}
    \label{fig:double_glitch_strength}
\end{figure}

\noindent  {\bf Effectiveness with multiple faults.} The protection, however, assumes that the fault affects at most one of the two randomized instances. As shown in Listing~\ref{lst:rsr_memcpy}, one instance of double fault that can bypass the detection mechanism is if both instances are corrupted to produce the same output~(targeting lines 3,8 resulting in zeroizing register {\tt r5}). In this case, the comparison ${\tt r6=r8}$ succeeds despite both values being faulty. Another instance targeting any instruction in one of the randomized computations and the {\tt cmp} instruction would also be sufficient to bypass the countermeasure. To practically validate this using EMFI, we observed that the attack requires two sharp glitches within an offset-window of $400$\;ns, which is challenging for our current setup, as the EM pulse needs at least $800$\;ns before regaining the required glitch strength. Figure~\ref{fig:double_glitch_strength} depicts the strength of the EM pulse spanned at an interval of $300$\;ns and $800$\;ns respectively. However, the ChipSHOUTER\footnote{\url{https://manuals.plus/m/119f1672d503d9743c5fd64815e7020386fc7f18539aa8015fb613c46607422f}} allows a minimum consecutive pulse spacing of 10\;ns, making it more suitable for evaluation. 

Even if an attacker has the capacity or specialized setup to perform consecutive glitches, our initial threat model assumptions remain; characterizing the target allows the identification of the multiple target points, thereby bypassing RSR with the necessary number of glitches. An adaptive version of RSR can require more than 2 glitches, thus increasing the overhead needed for the attacker to characterize and glitch the necessary points. However, implementing RSR introduces overhead as well, such as RSR applied to the \texttt{smlad}~(see Section \ref{rsr-smlar}).

\begin{lstlisting}[float=tb, captionpos=b, caption={Assembly code snippet illustrating RSR for \texttt{memcpy}}, label={lst:rsr_memcpy}]
# Randomized instance 1
ldrb     r4, [r1]
ldrb     r5, [rand1]
add      r6, r4, r5  # r6: = x + rand1
sub      r6, r6, r5  # r6: = x
# Randomized instance 2
ldrb     r7, [r1]
ldrb     r5, [rand2]
add      r8, r7, r5  # r8: = x + rand2
sub      r8, r8, r5  # r8: = x

cmp      r6, r8
bne      fault_detected

strb     r6, [r0]    # store result
ldrb     r9, [r0]    # read back destination
cmp      r9, r6
bne      fault_detected
\end{lstlisting}

\subsection{RSR Implementation for \texttt{smlad}} \label{rsr-smlar}
\noindent To protect \texttt{smlad}, we exploit its bilinear structure to construct two randomized instances that produce the same result in fault-free execution. For an \texttt{smlad} operation with operands $a$, $b$, and an initial accumulator $c$, we use 
\begin{equation*}
\texttt{SMLAD}(a,b,c) = \texttt{SMLAD}(a+r,b,c) - \texttt{SMLAD}(r,b,0)
\end{equation*}
where $r$ is a randomly generated packed operand. Listing~\ref{lst:rsr_smlad} shows the corresponding assembly-level implementation. Two independent random values, $r_1$ and $r_2$, are used to construct two RSR instances. Each instance computes the randomized \texttt{smlad} result and subtracts the corresponding random contribution to recover the original operation. The resulting values $y_1$ and $y_2$ are then compared using \texttt{cmp}. A mismatch triggers \texttt{fault\_detected}, indicating that the two randomized executions are inconsistent. The detection mechanism assumes that a fault affects at most one of the two randomized instances. A double fault can bypass the consistency check if both instances are corrupted such that they produce the same faulty output. Using simulation, we identified 8 instruction pairs that need to be skipped to successfully bypass the countermeasure. \smallskip

\noindent \textbf{RSR overhead.} RSR for the {\tt smlad} operation, while effective, incurs significant overhead. The primary computational overhead stems from the multiplications, thus slowing down each {\tt smlad} would almost directly line up with a slowdown of the overall running of the model. The instruction used in the example increases from one to ten, with the {\tt smlad} being a single cycle instruction, which decreases performance by an order of magnitude. Furthermore register usage doubles from four to eight, this being an optimized implementation where all registers are being used, would increase the amount of spilling to memory dramatically, decreasing performance further. 

\begin{lstlisting}[float=tb, captionpos=b, caption={Assembly code snippet illustrating RSR for \texttt{SMLAD} with a non-zero initial accumulator.}, label={lst:rsr_smlad}]
# Randomized instance 1  
# y_1 = SMLAD(a+r_1, b, c)-SMLAD(r_1, b, 0)
add     r6, r0, r3
smlad   r7, r6, r1, r2    # SMLAD(a+r_1, b, c)
smlad   r8, r3, r1, r5    # SMLAD(r_1, b, 0)
sub     r7, r7, r8        # y_1

# Randomized instance 2  
# y_2 = SMLAD(a+r_2, b, c)-SMLAD(r_2, b, 0)
add     r6, r0, r4
smlad   r8, r6, r1, r2    # SMLAD(a+r_2, b, c)
smlad   r9, r4, r1, r5    # SMLAD(r_2, b, 0)
sub     r8, r8, r9        # y_2

# Consistency check
CMP     r7, r8
BNE     fault_detected
MOV     r_result, r7  # r7 has verified result
\end{lstlisting}

\section{Limitations} 

\noindent \textbf{Multiple faults-triggered attack.}
We consider only single-fault triggers (one injected element). Backdoors that require multiple coordinated faults, e.g., sequential flips across locations or layers, are not characterized. 
While potentially stealthier, they compound the physical difficulty, since the per-attempt fault rates multiply and the injections must be tightly coordinated. 
We leave this direction to future work. \smallskip

\noindent \textbf{Characterization of more patterns.}
Our attack assumes the adversary has characterized the precise data-change pattern that the physical fault injection induces on the target.
In this paper, a single intermediate element is overwritten with a known value at a fixed buffer offset. 
We characterize only this restricted family of faults to showcase and ensure that the fault-induced changes are minimal enough to be stealthy.
Richer patterns have been observed in our experiments, such as multi-byte corruption, but these cases are not considered in this work. 
Extending the characterization to this broader space of fault patterns, and designing a corresponding backdoor to exploit it, is left to future work. \smallskip

\noindent \textbf{Attack setting with limited control of the target device.} To perform the attacks and investigate fault behavior we require sufficient control over the target device, such as the addition of trigger signals on the code level to signal when a target operation starts and ends. Adding trigger signals reduce the scanning time to find sensitive offset positions, while confirming that the desired target point is actually being attacked. If a trigger is absent, it will become more challenging for an adversary to perform the attack since identifying the target point from this black-box perspective is more challenging but not infeasible. Nonetheless, our attack remains a threat if an attacker manages to glitch at this target point, since the absence of triggers does not remove the vulnerability of a target point.  \smallskip

\noindent \textbf{FI reliability.} 
Regarding the number of faults, EMFI may require an attacker to perform multiple glitches before achieving a successful fault~\cite{moradi2011power}.
Following common practice in fault injection on cryptographic circuits~\cite{moradi2011power}, we assume the attacker can perform FI multiple times over a considerable time period, which shows the worst-case vulnerability. 
Regarding the reliable physical location, we find that it can be reproduced using similar parameter settings and a limited local search.
Our experiments show that a well-identified probe position can yield a larger FSR than other working positions, but the backdoor can be triggered with several trials at all working positions. 
Regarding reliable fault triggering in a practical physical environment, a pattern-based trigger generator~\cite{beckers2016design}, such as one based on the sum of absolute differences (SAD), can be used.
We leave the exploration of these methods to future work.
\smallskip

\section{Conclusion}\label{sec:conclusion}
\noindent This paper introduces \ourmethod: a backdoor attack that bridges algorithm-level attacks to implementation-level physical FI attacks. 
We propose the characterize-then-exploit method, which first characterizes the fault-induced data-level effects and then develops attacks that explicitly exploit them.
In particular, we connect FI with backdoor learning and propose cross-level attacks that jointly exploit implementation- and algorithm-level vulnerabilities.
Experiments targeting C-level and assembly-level implementations on ARM Cortex-M4 demonstrate that two different and inexpensive fault injection methods, i.e., EMFI and Voltage Glitching, can introduce precise faults that trigger backdoors without being easily detected by common backdoor detectors.
We highlight a novel cross-level attack vector at the intersection of hardware and algorithmic levels, stressing the need for defenses across abstraction layers.

\section*{Acknowledgment}
\noindent We thank Chris Berg from Keysight Technologies for their support with the physical inspection of the chip, and Péter Horváth for his help during the experimental setup. Furthermore, we thank all reviewers for their thorough and constructive feedback during the artifact and paper submission period. 
This work was supported by the European Union under Grant Agreement No. 101225691 QARC and by Dutch Research Council (NWO) through the following projects: PROACT (NWA.1215.18.014), TTW PREDATOR 19782, TTW BASES 20858, the CiCS project of the research programme Gravitation (024.006.037) and Open competition project ESCAPE (OCENW.XL.23.112).

\bibliographystyle{IEEEtran}
\bibliography{references}

\appendices

\section{FI Characterization on \texttt{memcpy}}
\label{appendix:memcpy}

\begin{figure}[!h]
    \centering
    \includegraphics[width=1\linewidth]{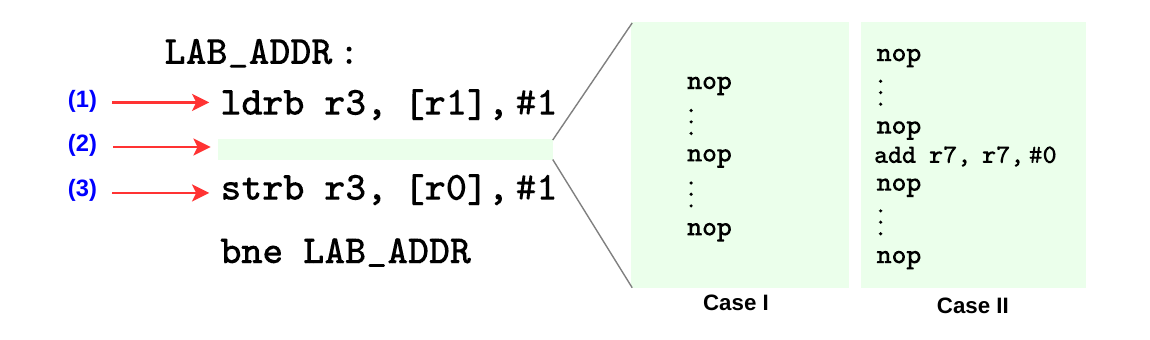}
    \caption{Different Attack Points in {\tt memcpy} operation. (1), (3) targets the load or store instruction to set register {\tt r3} to zero, respectively; (2) targets between the load and store instruction to set {\tt r3} to hold the value in {\tt r1} or {\tt r0}; {\bf Case I} introduces {\tt NOP} instructions before the target operation, whereas {\bf Case II} includes an add with 0 operation in addition to {\tt NOP} instructions. For target (1) and (3), we work with {\bf Case I}, and for target (2) we use {\bf Case I} and {\bf II}. }
    \label{fig:memcpy_ins}
\end{figure}

\noindent Figure~\ref{fig:memcpy_ins} depicts one iteration of {\tt memcpy} and the potential FI target points. We report our observations of glitching at the different target points as follows. The glitch parameters are summarized in Table~\ref{tab:emfi-transposed}.

\noindent \textbf{Glitching the \texttt{ldrb} or \texttt{strb} instruction. }
In normal execution, register \texttt{r3} fetches each pixel byte from \texttt{r1} sequentially. Thus before every fetch, \texttt{r3} holds the previous pixel value. 
However, in presence of FI, glitching {\tt ldrb} or {\tt strb} results in \texttt{r3} holding the value \texttt{0x00}. 
This is a consistent observation, confirming that the targeted \texttt{ldrb} or {\tt strb} is not skipped, rather the register content is corrupted.
A skipped instruction would rather result in {\tt r3} holding the previous pixel value.  

\noindent \textbf{Glitching between the \texttt{ldrb} and \texttt{strb}. }
We also glitch between the \texttt{ldrb} and \texttt{strb} instructions, as shown in Figure~\ref{fig:memcpy_ins}~(2). We consider two cases: i) inserting a sequence of {\tt NOPs} or ii) inserting {\tt NOPs} along with a functionally redundant operation~(adding zero to a register), to analyze the fault effects by comparing the register contents before and after the attack point.

In Case I, after the fault, \texttt{r3} holds the value \texttt{0xBF}~(see Figure~\ref{fig:emfiResultsMnistVisua0l}); which is the high byte encoding of a \texttt{NOP}.

To confirm the hypothesis that we indeed obtain the instruction encoding in {\tt r3}, we replace 1 NOP in the sequence with the instruction \texttt{orr r7, r7, r7} or \texttt{add r7, r7, \#0}~(referred to as Case II), and repeat FI multiple times. 
For most FI runs, we observe values \texttt{0x1d} or \texttt{0x64} in {\tt r3} while \texttt{0xBF} appeared occasionally. 
Comparing with the register contents post-FI, we identify that \texttt{r3} gets either the address stored in the destination register \texttt{r0}: \texttt{0x200420\textbf{64}} or the one in source register \texttt{r1}: \texttt{0x80303\textbf{1d}}. 
Since \texttt{strb} takes the lowest byte, we get \texttt{1d} or \texttt{64} as the resulting pixel byte. 
Notably, destination register address gets incremented, pointing to the next byte. Thus for each target pixel this value increases by 1 (e.g. byte 0 = \texttt{0x64}, byte 1 = \texttt{0x65}, byte 2 = \texttt{0x66} etc.), making it a predictable fault model given the starting memory address is known, and backdoor attacks can exploit it.

\begin{table}[h]
\caption{EMFI configurations targeting \texttt{memcpy}.}
\centering
\resizebox{\columnwidth}{!}{
\begin{tabular}{lccc}
\toprule
\textbf{Field} & \textbf{(1)} & \textbf{(2)} & \textbf{(3)} \\
\midrule
FI targets 
& {\tt ldrb}
& Between {\tt ldrb} \& {\tt strb} 
& {\tt strb} \\

Post-FI target byte value  
& 0x00 
& Tractable (0xBF, $\dots$) 
& 0x00 \\

Offset range (ns)  
& 44--50
& 68--70 
& 10 \\

Offset step size (ns) 
& 2 
& 2 
& N/A \\

Glitch Source Power (GSP) 
& 50\% 
& 58--60\% 
& 62\% \\

GSP step size 
& N/A 
& 1 
& N/A \\

\bottomrule
\end{tabular}
}
\label{tab:emfi-transposed}
\end{table}

\end{document}